\documentclass[12pt,letterpaper]{article}
\pdfoutput=1
\usepackage{jheppub2}
\usepackage{amsmath}
\usepackage{bm}
\usepackage{comment}
\usepackage{rotfloat}
\usepackage{amssymb}

\hypersetup{
    colorlinks=true,       
    linkcolor=red,          
    citecolor=blue,        
    filecolor=magenta,      
    urlcolor=blue           
}

\def\tr{\,{\rm tr}\,}

\title{Vacuum structure of Yang-Mills theory as a function of $\theta$}

\author[a]{Kyle Aitken,}
\author[b]{Aleksey Cherman,}
\author[c]{Mithat \"Unsal}
\emailAdd{kaitken17@gmail.com}
\emailAdd{aleksey.cherman.physics@gmail.com}
\emailAdd{unsal.mithat@gmail.com}
\affiliation[a]{Department of Physics, University of Washington, Seattle, WA 98105 USA}
\affiliation[b]{Institute for Nuclear Theory, University of Washington, Seattle, WA 98105 USA}
\affiliation[c]{Department of Physics, North Carolina State University, Raleigh, NC, USA}

\abstract{ 
It is believed that in $SU(N)$ Yang-Mills theory observables are $N$-branched functions of the topological $\theta$ angle.  This is supposed to be due to the existence of a set of locally-stable candidate vacua, which compete for global stability as a function of $\theta$.  We study the number of $\theta$ vacua, their interpretation, and their stability properties using  systematic semiclassical analysis in the context of adiabatic circle compactification on $\mathbb{R}^3 \times S^1$.  We find that while observables are indeed N-branched functions of $\theta$, there are only $\approx N/2$ locally-stable candidate vacua for any given $\theta$.  We point out that the different $\theta$ vacua are distinguished by the expectation values of certain magnetic line operators that carry non-zero GNO charge but zero 't Hooft charge.  Finally, we show that in the regime of validity of our analysis YM theory has spinodal points as a function of $\theta$, and gather evidence for the conjecture that these spinodal points are present even in the $\mathbb{R}^4$ limit.}

\preprint{INT-PUB-18-026}

\begin{document}

\maketitle

\section{Introduction}

Pure 4D $SU(N)$ Yang-Mills theory on a Euclidean spacetime manifold $M$ has a dimensionless $\theta$ parameter entering the action as
\begin{align}
S_{\theta} = \frac{i\theta}{8\pi^2} \int_M \tr F \wedge F
\end{align}
where $F$ is the $SU(N)$ field strength. Since the topological charge $Q = \frac{1}{8\pi^2} \int_M \tr F \wedge F$ is quantized, $Q \in \mathbb{Z}$, $\theta$ is a periodic parameter with period $2\pi$.  Hence one should expect that observables are periodic in $\theta$ with a period of $2\pi$.  However, considerations involving the large $N$ limit imply that the way that this periodicity is enforced is rather subtle \cite{DAdda:1978vbw,Witten:1978bc,Witten:1979vv,Witten:1980sp}.  Observables are not smooth $2\pi$-periodic functions of $\theta$.  Instead, they are expected to be multi-branched functions with non-analytic behavior at e.g. $\theta = \pi$.  The reason for this is that YM theory has many  locally-stable candidate vacuum states, and which one is the true globally-stable vacuum depends on $\theta$.  For instance, it is expected that the vacuum energy $E(\theta)$ takes the form
\begin{align}
E(\theta) = \min_{k} \tilde{E}_k(\theta/N)  \,,
\label{eq:vacuum_example}
\end{align}
where $\tilde{E}_k$ are the vacuum energies of the locally-stable candidate vacuum states labeled by an integer $k$.  The states $ \tilde{E}_k$ are expected to individually depend on $\theta/N$ and hence have a $2\pi N$ $\theta$ periodicity, and the $2\pi$ $\theta$ periodicity expected in $SU(N)$ YM theory is a consequence of the minimization in \eqref{eq:vacuum_example}.  These expectations have been verified in 2D and 4D models which are analogous to 4D pure YM, see e.g. Refs.~\cite{DAdda:1978vbw,Witten:1978bc,Witten:1979vv,Witten:1980sp,Witten:1998uka,Dine:2016sgq}.  Some aspects of the $\theta$ dependence of YM theory have also been extensively studied using numerical lattice simulations, see e.g.~\cite{DelDebbio:2002xa,Vicari:2008jw,Panagopoulos:2011rb,DElia:2012pvq,DElia:2013uaf,Bonati:2013tt,Bonati:2015vqz,Bonati:2016tvi}.

However, these standard observations leave open some interesting questions:
\begin{enumerate}
\item{What are the symmetries of YM theory as a function of $\theta$?}
\item{What is a physical interpretation of the different branches?  }
\item{How many candidate locally-stable vacuum branches does pure $SU(N)$ YM theory actually have?  }
\item{Does YM theory have spinodal points\footnote{By a spinodal point we mean a place where a metastable state reaches the limit of local stability.} as a function of $\theta$?}
\end{enumerate}

The answer to the first question above  is somewhat more subtle than is commonly assumed because center symmetry and charge conjugation and $PT$ symmetries do not commute. Explaining this is the focus of a companion paper \cite{Aitken:2018kky}.  In this paper, we focus on the other three questions, which  concern the dynamics of the theory.  Consequently, to answer them it is very helpful to consider a setting where the physics becomes amenable to analytic treatment.  In this paper we use the construction proposed in Ref.~\cite{Unsal:2008ch}, and intensively explored in related works, see e.g. Refs~
\cite{
Unsal:2007vu,Unsal:2007jx,Shifman:2008ja,Shifman:2008cx,Shifman:2009tp,
Cossu:2009sq,Myers:2009df,Simic:2010sv,Unsal:2010qh,Azeyanagi:2010ne,Vairinhos:2011gv,
Thomas:2011ee,Anber:2011gn,Unsal:2012zj,
    Poppitz:2012sw,
    Poppitz:2012nz,Argyres:2012ka,Argyres:2012vv,
    Anber:2013doa,Cossu:2013ora,
    Anber:2014lba,Bergner:2014dua,Bhoonah:2014gpa,Li:2014lza,
    Anber:2015kea,Anber:2015wha,Misumi:2014raa,
    Cherman:2016hcd,Aitken:2017ayq,Anber:2017rch,Anber:2017pak,Anber:2017tug,Anber:2017ezt,Ramamurti:2018evz}, where YM theory is compactified on a circle $\mathbb{R}^3\times S^1$ with stabilized center symmetry for all circle circumferences $L$.  The idea is that with center symmetry unbroken for all $L$, the physics is smooth in $L$, and becomes analytically calculable for small $L$.

First, we find that the different $\theta$ vacua turn out to be distinguished by the expectation values of certain magnetic line operators wrapping $S^1$.  Second, we give an explicit count of the vacua, with the result that there are only $\sim N/2$ locally-stable vacua for any given $\theta$. Nevertheless, observables are $N$-branched functions.\footnote{In the classic paper \cite{Witten:1980sp}, Witten asserted that at large-$N$, the number of metastable states is of {\it order N}, while observables are  exactly  $N$-branched functions.  These comments are of course consistent with our results, but are occasionally incorrectly interpreted to imply that there are necessarily exactly $N$ metastable branches in $SU(N)$ gauge theory.  }
  Third, we find that as $\theta$ is varied from $0$ to $2\pi N$, any given $\theta$ vacuum goes from being locally stable to being locally unstable. Hence the theory has spinodal points as a function of $\theta$ for any $N$. \footnote{Spinodal points in YM were conjectured to exist by Shifman \cite{Shifman:1998if} from some suggestive  extrapolations of softly-broken $\mathcal{N}=1$ SYM theory. The existence of spinodal points is also consistent with behavior seen in a holographic model of YM theory examined in \cite{Dubovsky:2011tu,Bigazzi:2015bna}, as well as with the behavior of QCD with light fundamental quarks \cite{Creutz:1995wf,Dubovsky:2010je,Smilga:1998dh}.  The novelty of our analysis is that it is done in a systematically-controlled setting \emph{without} any quark fields with masses $\lesssim \Delta(\theta = 0)$, where $\Delta(\theta = 0)$ is the mass gap at $\theta = 0$.} We also comment on the vacuum structure in various related models, and point out that the presence of spinodal points is sensitive to whether the theory contains light adjoint fermions.

   

The paper is outlined as follows. To keep the paper self-contained, we review some basic properties of circle-compactified YM theory with stabilized center symmetry in Sec.~\ref{sec:YM}, with an eye to how $\theta$ enters a small-$L$ 3D effective field theory description.  In Sec.~\ref{sec:interpretation_YM_vacua} we discuss the interpretation of $\theta$ vacua in both in the 3D EFT and from the 4D point of view.  Section~\ref{sec:condensates} discusses the behavior of parity-even and parity-odd gluon condensates.  Section~\ref{sec:genericity} then argues for the existence of spinodal points in center-stabilized YM theory and compares such findings to the behavior of a variety of other theories, with an eye to understanding the conditions under which one should expect the existence of spinodal points as a function of $\theta$.  Our results are summarized in Sec.~\ref{sec:outlook}.

\section{Center-stabilized YM theory on a small circle}
\label{sec:YM}

 Our main interest is in the $\theta$ dependence of the physics on $\mathbb{R}^4$.  Since this is a strongly-coupled limit of the theory, we will have to be satisfied with studying the $\theta$ dependence in some calculable corner of the phase diagram which is smoothly connected to $\mathbb{R}^4$.   This calculable corner of the phase diagram appears when one studies the theory on $\mathbb{R}^3\times S^1$.

$SU(N)$ YM theory has a global $\mathbb{Z}_N$ center symmetry, and the long distance physics is highly sensitive to its realization, which can depend on $L$.   Center symmetry is known to be automatically preserved at large $L$ in pure YM theory.  Thus to have any chance that a small-$L$ limit would be smoothly connected to large $L$, center symmetry must also be preserved for small $L$.  Fortunately, it is known that center symmetry can be stabilized at small $L$ by  certain double-trace deformation, or by adding heavy adjoint fermions to the theory \cite{Unsal:2008ch,Azeyanagi:2010ne,Unsal:2010qh}. A large amount of evidence \cite{Unsal:2008ch,
Unsal:2007vu,Unsal:2007jx,Shifman:2008ja,Shifman:2008cx,Shifman:2009tp,
Cossu:2009sq,Myers:2009df,Simic:2010sv,Unsal:2010qh,Azeyanagi:2010ne,Vairinhos:2011gv,
Thomas:2011ee,Anber:2011gn,Unsal:2012zj,
    Poppitz:2012sw,
    Poppitz:2012nz,Argyres:2012ka,Argyres:2012vv,
    Anber:2013doa,Cossu:2013ora,
    Anber:2014lba,Bergner:2014dua,Bhoonah:2014gpa,Li:2014lza,
    Anber:2015kea,Anber:2015wha,Misumi:2014raa,
    Cherman:2016hcd,Aitken:2017ayq,Anber:2017rch,Anber:2017pak,Anber:2017tug,Anber:2017ezt} suggests that with this setup, the physics depends smoothly on $NL\Lambda$, without phase transitions or rapid crossovers.  (The reason why $NL\Lambda$ rather than $L\Lambda$ is the relevant dimensionless parameter will be reviewed below.)  When $NL\Lambda$ is large, the physics approaches that of pure YM on $\mathbb{R}^4$, with strong coupling at large distances.  But when $NL\Lambda$ is small, the physics remains weakly coupled at all distances.  This enables  systematic semiclassical studies of the non-perturbative dynamics, with $NL\Lambda$ interpreted as the small expansion parameter. 
    
Let us denote the compactified direction by $x_4$, and label the non-compact directions $x^{\mu}, \mu = 1,2,3$.  In the presence of the center-stabilizing double-trace deformation, the holonomy $\Omega = P \exp (i \oint dx_4 A_4)$ gets a $\mathbb{Z}_N$-symmetric vacuum expectation value (that is, with vanishing expectation values for traces of $\Omega^n$ for $n \neq n \textrm{ mod } N$) for all $L$.  At small $NL\Lambda$, where the theory is weakly coupled, the eigenvalues of $\Omega$ take the values
\begin{align}
\Omega = \omega^{-(N-1)/2} \mathrm{diag}(1,\omega, \ldots, \omega^{N-1}),\qquad \omega = e^{2\pi i /N}
\label{eq:center_sym_holonomy}
\end{align}
up to permutations generated by the Weyl group $S_N \subset SU(N)$.  
From the perspective of the 3D effective field theory, valid at long distances compared to $L$, $\langle \tr \Omega^n \rangle = 0 $ for $n \neq 0$ (mod $N$)   implies (in the usual gauge-fixed sense) that  compact $\langle A_4 \rangle \neq 0$, and $A_4$  acts as an adjoint Higgs field ``breaking'' the $SU(N)$ gauge group down to $U(1)^{N-1}$. The color-off-diagonal components of the gauge fields $A_{\mu}$, where now and henceforth $\mu = 1,2,3$, pick up effective masses which are integer multiples of
\begin{align}
m_W \equiv \frac{2 \pi}{ NL}.
\end{align}
Indeed, this is why the theory becomes weakly coupled at long distances so long as $m_W \gg \Lambda$:  the gauge coupling stops running at the energy scale $m_W$, since there are no charged modes below $m_W$.

The light degrees of freedom at small $L$ are the $N-1$ Cartan gluons, with field strengths we denote as $F_{\mu \nu}^{a}, a = 1, \ldots N-1$.  The Cartan gluons will often be referred to as ``photons".  Even though it is rather natural to interpret the index `a' as a color index, it is important to note that this index has a manifestly gauge-invariant meaning.  In particular, one can show that
\begin{align}
\label{eq:gi_famu}
F^{a}_{\mu \nu}(x_{\mu})  \sim \frac{1}{N} \frac{1}{L} \int d x_4 \sum_{q=1}^{N-1} \omega^{-q a} \tr \Omega^{q}(x_4, x^{\mu}) F_{\mu \nu}(x_4,x_{\mu}) \, ,
\end{align}
where $F_{\mu \nu}$ is the 3D part of the full non-Abelian $SU(N)$ field strength, $\Omega(x_4,x_{\mu})  =   \Omega(x_{\mu}) = \mathcal{P} \exp{\left[i \int_{x_4}^{L+x_4} d x^{\prime}_4 A_4(x^{\prime_4}, x_{\mu}) \right]}$.  We use a $\sim$ in relating the left and right hand sides of Eq.~\eqref{eq:gi_famu} to emphasize that the right-hand-side should be viewed as a 4D manifestly gauge-invariant interpolating operator for the expression on the left-hand side.  That is, at weak coupling, the 4D operator on the right dominantly couples to single-3D-``photon'' states, but it also creates e.g. pairs of W-bosons with the same quantum numbers.  

For the purposes of later discussion, it will be notationally convenient to introduce an fictitious $N$-th photon, $F_{\mu \nu}^{N}$, which will decouple from the $N-1$ physical photons both perturbatively and non-perturbatively.
With this done,  the photon action can be written as
\begin{align}\label{eq:euc_action}
S_{\rm tree}  = \frac{L}{4g^2} \int d^3{x} \sum_{a =1}^{N} F^{a}_{\mu \nu} F^{a\mu \nu} \,.
\end{align}
Equation \eqref{eq:gi_famu} can be used to infer that under center symmetry, the Cartan gluons transform as 
\begin{align}
\mathcal{S}: \; F^{a}_{\mu \nu} \to F^{a+1}_{\mu \nu} \,. 
\label{eq:center3d}
\end{align}

The Cartan gluons are gapless to all orders in perturbation theory, see e.g.~\cite{Davies:1999uw,Davies:2000nw,Unsal:2008ch}, but develop a mass non-perturbatively.  To see this, we pass to the Abelian dual representation of the effective action for dual photons.  At tree level, the effective action becomes
\begin{align}
S_{\rm tree, dual}  = \lambda m_W \int d^3{x}  \sum_{a =1}^{N} (\partial_{\mu}\sigma^a)(\partial^{\mu}\sigma^a) \equiv \lambda m_W \int d^3{x} ~ (\partial_{\mu}\vec{\sigma})^2.
\label{eq:perturbative_action}
\end{align}
Here $\sigma^a$ are the dual photons, $\vec{\sigma} = (\sigma^1, \ldots, \sigma^N)$, and the $\sigma^a$ fields are related to the original Abelian field strengths by the 3d Abelian duality relation 
\begin{align}
F_{\mu \nu}^{a} = \lambda m_W/(4\pi^2) \epsilon_{\mu\nu\rho} \partial^{\rho} \sigma^{a} \, .
\end{align}
The tree-level dual photon effective action in \eqref{eq:perturbative_action} gives rise to a conserved shift current $J^a_{\mu} = \partial_{\mu}\sigma^a$.  The symmetry conservation equation $\partial^{\mu}J_{\mu}^a = 0$ can be rewritten as $\partial^{\mu} \epsilon_{\mu \alpha \beta} \vec{F}^{\alpha \beta} = 0$.  
The latter equation always holds in the absence of magnetic monopoles, and the theory we are studying certainly has no magnetic monopole field configurations in perturbation theory.  So the $\vec{\sigma}$ shift symmetry is exact in perturbation theory, and the dual photons are perturbatively gapless.%
\footnote{Perturbative interactions do not generate a potential for $\vec{\sigma}$, but they do generate a non-trivial metric on the $T^{N-1}/S_N$ target space, $ \sum_{a =1}^{N} (\partial_{\mu}\sigma^a)(\partial^{\mu}\sigma^a) \to  \sum_{a,b =1}^{N} f_{a,b}(\lambda)(\partial_{\mu}\sigma^a)(\partial^{\mu}\sigma^b)$, where $f_{a,b} = \delta_{a,b} + \mathcal{O}(\lambda)$.  The implications of this fact were discussed in detail in \cite{Anber:2014lba}, with some large $N$ implications covered in \cite{Cherman:2016jtu}.  Here we take $f_{a,b} = \delta_{a,b}$ since the $\mathcal{O}(\lambda)$ corrections do not materially affect our discussion.}%

We now briefly review the leading non-perturbative effects. More details of such solutions can be found in, for example, ref. \cite{Anber:2011de}.  As in any $SU(N)$ YM theory, there exist instanton solutions with action $S_I = 8\pi^2 N/\lambda$ and topological charge $Q = 1$.  But in the setting of \eqref{eq:center_sym_holonomy}, the instantons  fractionalize into $N$ types of monopole-instantons \cite{Lee:1997vp,Kraan:1998pm}, each carrying topological charge $Q = 1/N$.   The monopole-instantons are the field configurations with the smallest finite value of the action.   Their actions are all equal, $S_0 = 8\pi/\lambda$, so that $N S_0$ coincides with the BPST instanton action $S_I$.  The magnetic charges of $N-1$ of these monopole-instantons are given by the $SU\left(N\right)$ \mbox{(co-)root} vectors,  $\vec{\alpha}_{a}, a=1, \ldots N-1$, and would be present even in a locally-3D theory.   The $N$-th monopole (the KK monopole) with action $S_0$ has magnetic charge given by the affine \mbox{(co-)root} $\vec{\alpha}_{N} = -\sum_{a}^{N-1} \vec{\alpha}_a$, and is present due to the locally four-dimensional nature of our theory\cite{Lee:1997vp,Kraan:1998pm}. \footnote{We choose to use an $N$-dimensional basis for the root vectors, so that their components can be written as $(\vec{\alpha}_a)_{i} = \delta_{a,i} - \delta_{a,i+1}$. All of these vectors are are orthogonal to $\vec{e}_0 = (1,1, \cdots, 1)$, $\vec{\alpha}_a \cdot \vec{e}_0 = 0$, where 
 $\vec{e}_0$ is the magnetic charge vector associated with the fictitious $N$-th photon.  This ensures that the fictitious photon completely decouples  from the physical fields.} More precisely, in terms of the $U(1)^N$ valued $3d$ magentic field defined by $\vec{B}^{\mu} \equiv \frac{1}{2g} \epsilon^{\mu\nu\rho 4} \vec{F}_{\nu\rho}$, the monopole of type $a$ has magnetic charge which satisfies 
\begin{align}\label{eq:monopole charge}
\int_{S^2} d\Sigma_{\mu} \vec{B}^{\mu} = \frac{2\pi}{g}\vec{\alpha}_a
\end{align}
with $S^2$ a sphere in $\mathbb{R}^3$ surrounding the monopole. The gauge invariance of \eqref{eq:monopole charge} can be made manifest by rewriting $\vec{F}_{\mu\nu}$  using \eqref{eq:gi_famu}. Since the monopole-instanton field configurations carry magnetic charge, their contributions to the path integral can produce a mass gap for the $\sigma^{a}$ fields.

At leading order in the semi-classical approximation, the magnetic monopole-instanton contribution to the effective action can be evaluated using the dilute monopole-instanton gas approximation. This approximation is under systematic control because the monopoles have a fixed characteristic size $\sim m_W^{-1}$, while their typical Euclidean separation is exponentially large, $\sim m_W^{-1} e^{+S_0/3}$. This gives
\begin{align}
S_{\rm \vec{\sigma}} = \int d^3{x}\, \left[ \lambda m_W (\partial_{\mu} \vec{\sigma})^2  + V(\vec{\sigma}) \right]
\label{eq:YM_EFT}
\end{align}
where 
\begin{equation}
V\left(\vec{\sigma}\right)=V^{(1)} + V^{(2)} + \ldots \, .
\end{equation}
where the expansion is in powers of $e^{-S_0}$, so that 
\begin{align}
V^{(1)} &= -\frac{A}{\lambda^{2}}m_{W}^{3}e^{-S_0}\sum_{a=1}^{N}\cos\left[\vec{\alpha}_{a}\cdot\vec{\sigma}+\frac{\theta}{N}\right]
\label{eq:YM_potential} \\
V^{(2)} &\sim e^{-2S_0} \sum_{a=1}^{N} \cos\left[ (\vec{\alpha}_{a} -\vec{\alpha}_{a+1})\cdot \vec{\sigma} \right] 
+ \cdots
\label{eq:magnetic_bion_contribution} 
\end{align}
Here $A$ is an $\mathcal{O}\left(1\right)$ scheme-dependent constant, while $V_2 \sim e^{-2S_0}$ represents the contributions arising at the next (second) order in the semiclassical expansion in powers of $e^{-S_0}$.  Starting from the second order in the semiclassical expansion, the $\sigma$ potential receives contributions from correlated monopole-instanton events.  The most important of these for our story are  \emph{magnetic bions}, which carry magnetic charge but have vanishing topological charge \cite{Unsal:2007jx}, and their contribution is written explicitly above.
At  second order in the semiclassical expansion one also encounters \emph{topological bions}, which carry both magnetic and topological charges, and \emph{neutral bions}, which have neither topological nor magnetic charge.  The neutral bions do not contribute to the $\sigma$ potential.  The topological bions do contribute, but their contributions are suppressed by positive powers of $\lambda$ relative to the magnetic bion contribution if the center symmetry is stabilized by massive adjoint fermions or typical double-trace deformations \cite{Unsal:2012zj}, and their effects are subsumed in the $\ldots$ in Eq.~\eqref{eq:magnetic_bion_contribution}. 
The magnetic bions can become important if the leading-order contribution vanishes, which can happen for  certain values of $\theta$ and $N$.  

The potential $V^{(1)}$ has $N$ extrema within a unit cell of the weight lattice given by
\begin{equation}
\vec{\sigma}_{k}=\frac{2\pi k}{N}\vec{\rho},  \qquad \mathrm{with}\qquad\vec{\rho}\equiv\sum_{i=1}^{N-1}  \vec{\mu}_{i}
\label{eq:YMextrema}
\end{equation}
where $k=0,1,\ldots,N-1$, $\vec{\rho}$ is the Weyl vector, and $\vec{\mu}_i$ are the fundamental root vectors of $SU(N)$. These vectors satisfy $\vec{\alpha}_i \cdot \vec{\mu}_j = \delta_{ij}, \vec{\alpha}_i\cdot\vec{\rho}=1$ for $i=1,\ldots,N-1$, and $\vec{\alpha}_N\cdot\vec{\rho}=1-N$, which we will use below. Occasionally, it  will be useful to use an explicit basis where the root vectors are $(\alpha_a)_b=\delta_{a,b}-\delta_{a+1,b}$, $1\leq a <N$, $\vec{\sigma}_k$, in which case the extrema take the form
\begin{align}
\vec{\sigma}_{k} = \frac{2\pi k}{N} (N, N-1, \ldots, 2,1) \, .
\label{eq:weyl_vector}
\end{align}
The value of $V^{(1)}$ evaluated at each of the extrema is given by
\begin{equation}
V_{k} \equiv V\left(\vec{\sigma} = \vec{\sigma}_k \right)=-N\frac{A}{\lambda^{2}}m_{W}^{3}e^{-S_0}\cos\left(\frac{2\pi k+\theta}{N}\right) + \mathcal{O}(e^{-2S_0}) \,.
\label{eq:pot_vals_tnm}
\end{equation}
In the following sections we explore the vacuum structure and $\theta$ dependence of observables that follow from this effective field theory description.

It is also interesting to explore the transformations of these extrema under various symmetries of the system (such as center symmetry, parity, and charge conjugation symmetries) as a function of $\theta$.  We do so in a companion paper \cite{Aitken:2018kky}.  This analysis shows that center symmetry and e.g. charge conjugation do not commute, and as a consequence the group of discrete symmetries generically involves the dihedral group $D_{2N}$.  However, at $\theta=\pi$, the symmetry group becomes centrally extended and involves $D_{4N}$, which is consistent with  the results of Ref.~\cite{Gaiotto:2017yup} concerning mixed 't Hooft anomalies involving center symmetry and $CP$ symmetry.

\section{Vacuum structure of Yang-Mills theory as a function of $\theta$}
\label{sec:vac_YM}

\subsection{Spectrum and ground state properties}
We now examine the spectrum as a function of $\theta$.   The value of the leading-order potential at its extrema is given in Eq.~\eqref{eq:pot_vals_tnm}.  Diagonalizing the matrix of quadratic fluctuations around each extremum  tells us that the ``mass-squares" of fluctuations around the extrema are 
\begin{equation}
m_{q,k}^2 =m_{\gamma}^{2}\sin^{2}\left(\frac{\pi q}{N}\right)\cos\left(\frac{2\pi k+\theta}{N}\right) + \mathcal{O}(e^{-2S_0}) \, ,
\label{eq:extremum_spectrum}
\end{equation}
where $m_{\gamma}^2 \sim m_W^2 e^{-S_0}$ and $q=1,\ldots,N-1$. This basic formula has been known for a long time, see e.g. Refs.~\cite{Thomas:2011ee,Unsal:2012zj,Bhoonah:2014gpa}.  Here we give a precise count of the number of locally-stable vacua, and comment on the important role of magnetic bion corrections given by Eq.~\eqref{eq:magnetic_bion_contribution}. 

If all fluctuations have positive mass squares, a given extremum is a local minimum and hence corresponds to an (at least) metastable vacuum. It is straightforward to verify that $m_{q,k}^2 >0$ for all $q$ so long as the energy density $V_{k}$ is positive.  The true vacuum, corresponding to the global minimum, has the smallest value of $V_{k}$ among the local minima. For $-\pi<\theta<\pi$ the true vacuum is $k=0$, while for $(2M-1)\pi<\theta<(2M+1)\pi$ with $M\in\mathbb{Z}$ it is $k=N-M$ mod $N$.
 
Which vacua are locally stable also depends on $\theta$ and $N$.  For example, for $\theta = 0$ and $N=5$, $k = 0,1,4$  are local minima, but $k=2,3$ are local maxima, while for $\theta = \pi$ and $N=5$, $k = 0,4$ are local minima while $k=1,2,3$ are local maxima.   In general, for any $\theta$, roughly $N/2$ extrema of Eq.~\eqref{eq:YM_potential} are locally stable, while $N/2$ extrema are locally unstable. More precisely, we find that the number of locally stable vacua, $N_s$, is
\begin{align}
\label{eq:vacuum_count}
N_s = \begin{cases}
1+2 \lfloor \frac{N}{4} \rfloor & \theta = 0 \\
\lfloor \frac{N}{4} \rfloor +\lfloor \frac{N+3}{4} \rfloor  & 0<\theta < \pi/2 \\
\lfloor \frac{N + 1}{2}\rfloor & \theta = \pi/2 \\
\lfloor \frac{N+1}{4} \rfloor+\lfloor \frac{N+2}{4} \rfloor & \pi/2 < \theta < \pi \\
2 \lfloor \frac{N+2}{4} \rfloor &   \theta = \pi. 
\end{cases}
\end{align}
where the floor function $\lfloor x \rfloor$ gives the largest integer less than $x$.

It is amusing to note that for some values of $N$, there are values of $k$ such that the mass matrix of Eq.~\eqref{eq:extremum_spectrum} can vanish identically at leading order in the semiclassical expansion.  
But in the small $NL\Lambda$ regime, the subleading contribution to the masses comes from magnetic bions, and is always positive.\footnote{The heuristic argument for this involves two steps.  First, one notes that the magnetic bion contribution does not depend on $\theta$ since magnetic bions have zero topological charge.  
Second, there are examples of theories, such as adjoint QCD\cite{Unsal:2007jx} where the presence of fermion zero modes eliminates the monopole-instanton contribution to the $\vec{\sigma}$ potential, so the magnetic-bion contributions are leading order.  
Therefore, the bion contribution must by itself yield a positive mass spectrum for stability of the theory. }  
 So, when $k$ and $N$ conspire to make  the leading order masses vanish, taking subleading contributions into account, such as $V_2$ in Eq.~\eqref{eq:YM_potential}, implies that these extrema are in fact metastable vacua.  For example, when $N = 4$, and $\theta = 0$, the global minimum corresponds to $k=0$.  But if we set $k=1$ with $\theta = 0$, then $m_{q,1} = 0$ to leading order in the semiclassical expansion \cite{Bhoonah:2014gpa}.  
But the magnetic bion contribution to the mass  in Eq.~\eqref{eq:magnetic_bion_contribution} is always positive for any $N$ and $\theta$, so in fact the dual photon  mass spectrum at $k=1$ is
\begin{align}
m^2_{q,1} =  e^{-2S_0} \sin^4\left(\frac{\pi q}{4}\right) + \mathcal{O}\left(e^{-3S_0}\right) \,, \qquad N=4
\end{align}
So the $k=1$ extremum is in fact a local minimum for $N=4, \theta = 0$, and corresponds to a metastable state of the system.  These comments generalize to any $N$, and the locally-stable-vacuum counting formulas in Eq.~\eqref{eq:vacuum_count} take into account the effects of magnetic bions.

These results imply that the $2\pi$ periodicity of the system in its thermodynamic ground state is \emph{not}  enforced by a dance between $N$ candidate ground states which are all metastable, as is sometimes assumed in the literature. Instead, as $\theta$ is varied between $0$ and $2\pi N$, any given extremum passes from being locally stable to locally unstable, and for any given $\theta$, only $\sim  N/2 $ extrema are metastable vacua.\footnote{For $N=3$ we have explicitly verified that as an extremum passes from being a minimum to being a maximum, it merges with a saddle-point of the potential in Eq.~\eqref{eq:YM_potential}.  We expect this to generalize to all $N>3$.  The $N=2$ case must be treated separately, since the leading-order potential strictly vanishes at $\theta = \pi$, and one must take into account the magnetic bion contributions.  With this done, one finds that the spinodal point is associated with a merger of extrema of the leading-order potential with some extra extrema of the magnetic-bion-corrected potential. }

Lastly, we note that the 4D energy density $\mathcal{E}_k \equiv V_k/L$ is  
\begin{align}
\mathcal{E}_k \sim N^2 \Lambda^4 (NL\Lambda)^{-1/3} \cos\left(\tfrac{2\pi k +\theta}{N}\right).
\end{align}
On $\mathbb{R}^4$, it is expected that $\mathcal{E}_k(\theta = 0) \sim N^2$, while the topological susceptibility 
$\chi_{\rm top}$
\begin{align}
 \chi_{\rm top} \equiv \left\langle \frac{\partial^2 \mathcal{E}_k }{\partial\theta^2} \right\rangle \bigg|_{\theta = 0} \sim \mathcal{O}(N^0)
\end{align}
To get a calculable large $N$ limit where our formulas apply, one must take the double-scaling limit $NL\Lambda \sim N^0$ as $N\to \infty$ \cite{Unsal:2008ch,Cherman:2016jtu}, and ensure that $NL\Lambda \ll 1$.  In this limit, the $N$ scaling of $\mathcal{E}(\theta = 0)$ and of $\chi_{\rm top}$ is completely consistent with the expectations above.

\subsection{Interpretation of $\theta$-vacua}
\label{sec:interpretation_YM_vacua}

We have found that the vacua of YM theory on $\mathbb{R}^3\times S^1$ are labeled by the values of $\langle \tr \Omega^n \rangle, n \neq 0\,\, \mathrm{mod}\,\, N$ and  $\vec{\sigma}_k$. Here we discuss the interpretation of the physics behind the distinct extrema labeled by $k$ where $\langle\vec{\sigma}\rangle=\vec{\sigma}_k$, defined by \eqref{eq:YMextrema}.  We note that Ref.~\cite{Anber:2017rch} has  insightful remarks of a similar spirit to some of our discussion below.

Before we start, it will be helpful to recall some properties of the monopole-instanton operators.  In our dualized language, a type $a$ monopole of charge $\vec{\alpha}_a$ located at position $x^\mu$ is equivalent to the insertion of $e^{i\vec{\alpha}_a\cdot \vec{\sigma}(x^\mu)}$ in the path integral. In terms of the $SU(N)$ field strength, the magnetic charge of such a field configuration is defined by \eqref{eq:monopole charge}. Our discussion above implies that the $\theta$ vacua are all confining and labeled by $\theta$-dependent phases for these operators, so
\begin{align}
\langle \mathcal{M}_a \sim e^{-8\pi^2/\lambda} e^{i \vec{\alpha} \cdot \vec{\sigma}} \rangle =  e^{-8\pi^2/\lambda} e^{2\pi i k/N} , \qquad \langle \tr \Omega^n \rangle = 0\;,
\end{align}
where $n \neq 0 \, \mathrm{mod}\, N$ and the integer $k$ depends on $\theta$.

Let us start with comments on the physical interpretation of the $\theta$-extrema within the  3D effective theory. From our discussion in the previous paragraph, it is clear that $\langle\vec{\sigma}\rangle=\vec{\sigma}_k$ can be interpreted as defining uniform background distributions of magnetic monopoles of types $1$ to $N-1$ of with fractional magnetic charges $\frac{2\pi k}{N}\vec{\alpha}_{i}$, as well as the contribution from monopole of type $N$, giving a charge $\frac{2\pi k}{N}\left(1-N\right)\vec{\alpha}_{N}$.   One might be concerned that this would cause the system to have a divergent magnetic Coulomb energy, but fortunately this is not the case.  For arbitrary $k$, a background of constant $\vec{\sigma}=\frac{2\pi k}{N}\vec{\rho}$ represents magnetic charge
\begin{equation}
\sum_{i=1}^{N-1}\frac{2\pi k}{N}\vec{\alpha}_{i}+\frac{2\pi k}{N}\left(1-N\right)\vec{\alpha}_{N}= -2\pi k \, \vec{\alpha}_{N}
\end{equation}
with respect to the $U(1)^N$ gauge fields. This is equivalent to $\left\langle \vec{\sigma} \right\rangle=0$ for integer $k$ due to the $2\pi$ periodicity of $\vec{\sigma}$.

\begin{figure}[t]
\begin{center}
\includegraphics[width = .7\textwidth]{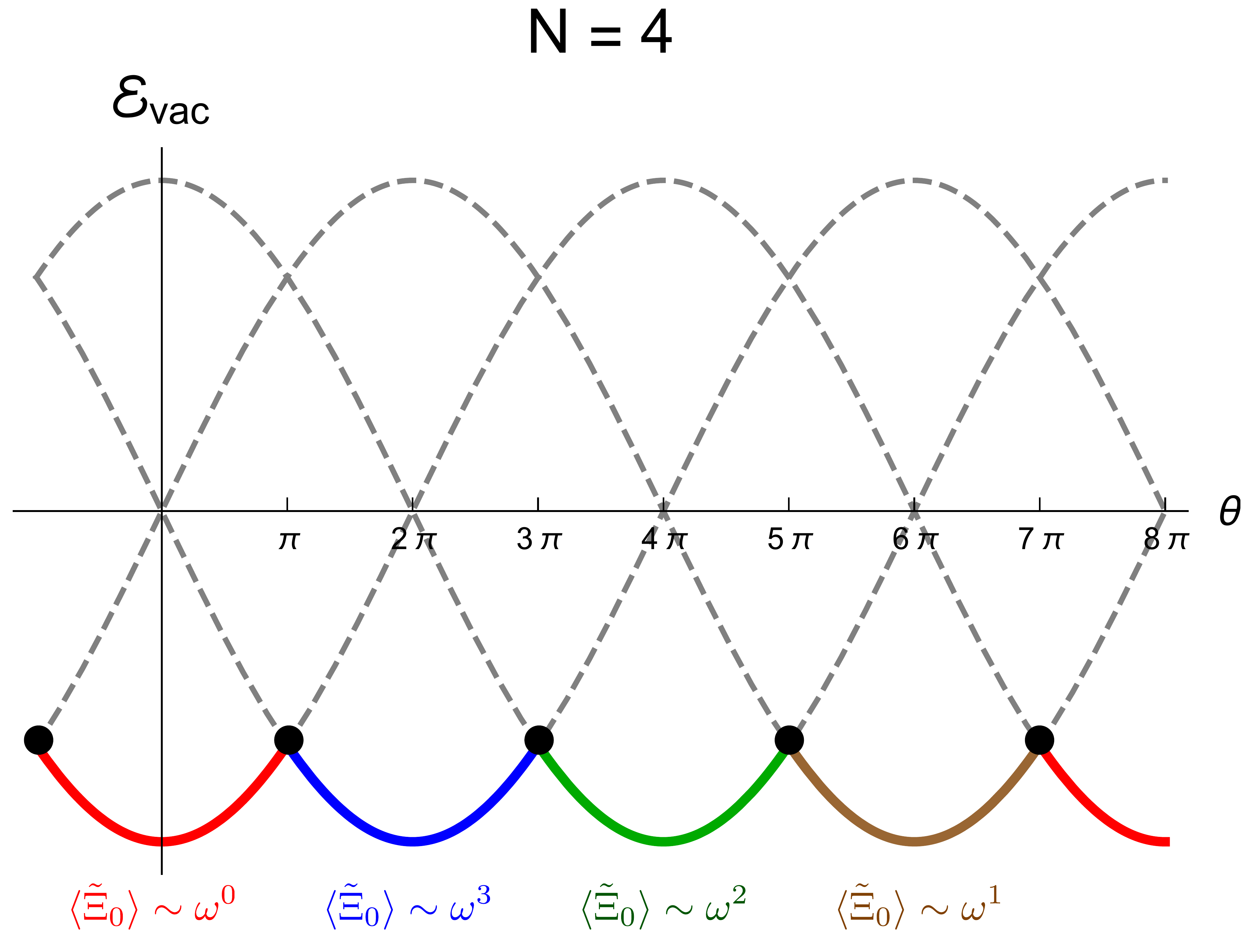}
\caption{[Color Online.] An sketch of the phase structure of $SU(N)$ YM theory for $N=4$ as a function of $\theta$.  The vertical axis depicts the vacuum energy density.  The gray dashed curves indicate the energy density associated with the four distinct $\theta$-extrema. The colored bold curves indicate the vacuum energy density associated with the thermodynamically stable vacuum branch for any given $\theta$.  The black dots at the cusps label the locations of quantum phase transitions.  These phase transitions are associated with changes in the expectation values of GNO 't Hooft magnetic holonomies, as discussed in the main text. }
\label{fig:vacuaYM}
\end{center}
\end{figure}

We now ask about the interpretation of the vacuum structure from the point of view of the four-dimensional $SU(N)$ gauge theory. To get such an interpretation, we need to better understand the meaning of operators like $e^{i \vec{\alpha}_a \cdot \vec{\sigma}(x^\mu)}$ in a 4D language. Thanks to Eq.~\eqref{eq:gi_famu}, we can write a manifestly gauge-invariant 4D interpolating operator for $\sigma_{a}$ as a line integral over the $S^1$ direction:
\begin{align}
\sigma_{a}(x^\mu) &= \frac{4\pi^2}{\lambda m_W} \frac{1}{\nabla^2}\left[\partial_{\rho} \epsilon^{\mu\nu\rho}F_{\mu \nu}^{a}\right] \\
&= \frac{2\pi}{\lambda m_W^2} \frac{1}{\nabla^2} \left[\partial_{\rho} \epsilon^{\mu\nu\rho} \int d x_4 \sum_{q=1}^{N-1} \omega^{-q a} \tr \Omega^{q}( x_{\mu}) F_{\mu \nu}(x_4,x_{\mu})\right] \,.
\end{align}
This can then be used to write a gauge-invariant 4D interpolating operator for $e^{i \vec{\alpha}_a \cdot \vec{\sigma}(x)}$, for which we will introduce a new symbol $\Xi_a$,
\begin{align}
\Xi_a(x_\mu) \sim \exp\left(\frac{2\pi i}{\lambda m_W^2} \frac{1}{\nabla^2} \left[\partial_{\rho} \epsilon^{\mu\nu\rho} \int d x_4 \sum_{q=1}^{N-1} (\omega^{-q a}-\omega^{-q (a+1)}) \tr \Omega^{q} F_{\mu \nu}\right]\right) \, .
\label{eq:4D_monopole}
\end{align}
This gauge-invariant expression makes clear that the index $a$ on $\mathcal{M}_a$, which at first glance may like a color label, is better interpreted as a discrete Fourier transform of center charge.  To make this clearer, one can define
\begin{align}
\tilde{\Xi}_{p} = \frac{1}{N} \sum_{a = 0}^{N-1} \omega^{a p} \Xi_{a}.
\end{align} 
The operators $\tilde{\Xi}_{p}, p = 1, \cdots N-1$ carry non-trivial center charge.  If they picked up non-trivial expectation values, this would be a signal of spontaneous center symmetry breaking.  But  on the $k$-th $\theta$-extremum one finds
\begin{align}
\langle \tilde{\Xi}_{0} \rangle \sim e^{2\pi i k/N}, \qquad   \langle \tilde{\Xi}_{p} \rangle  = 0 \,, p = 1, \ldots, N-1 \, .
\end{align}
So as the value of $k$ labeling the globally stable vacuum changes as a function of $\theta$, center symmetry is always unbroken, but the phase of $\langle \tilde{\Xi}_{0} \rangle$ changes in discrete steps.  This is illustrated in Fig.~\ref{fig:vacuaYM}.

To get some further insight into the interpretation of the $\Xi_a(x^\mu)$ operators, note that they are point-like in $\mathbb{R}^3$ and carry magnetic charge valued in the root
\footnote{Strictly speaking, the magnetic charges live in the root lattice of the dual magnetic group $G_L$ for  the `electric' gauge group  $G  = SU(N)$, which is $SU(N)/\mathbb{Z}_N$, and consequently the monopole operators should written as $e^{i \vec{\alpha}^{*}_a \cdot \vec{\sigma}}$, where $\vec{\alpha}^*_a$ are co-root vectors.  But for $G = SU(N)$ it happens that $\vec{\alpha}_a$ and $\vec{\alpha}^*_a$ are numerically identical, so we ignore this technicality in the main text.} 
 lattice of $SU(N)$, rather than in the fundamental weight lattice.  The magnetic charge in question is the one described in the classic work of Goddard, Nuyts, and Olive (GNO) \cite{Goddard:1976qe}.  Putting this together with Eq.~\eqref{eq:4D_monopole}, we interpret  $\Xi_a(x^\mu)$ as (GNO) magnetic line operators wrapping $S^1$. (This is consistent with the identification in  \cite{Argyres:2012ka}.) 

Since $\Xi_a(x^\mu)\sim e^{i \vec{\alpha}_a \cdot \vec{\sigma}(x^\mu)}$ is independent of the $x_4$ coordinate, it can be equivalently characterized by its properties in the $\mathbb{R}^3$ subspace at a given value of $x_4$ where it appears as a point-like object. In particular, consider some $S^2$ in the $\mathbb{R}^3$ subspace surrounding the point $x^{\mu}$ where we insert $\Xi_a$.  Then one can define $\Xi_a$ by demanding that it obey \eqref{eq:monopole charge}, with $F_{\mu\nu}^a$ replaced by its gauge-invariant interpolating operator \eqref{eq:gi_famu} if one wants to make gauge-invariance manifest. 
 
We note that 't Hooft proposed another sort of magnetic line operator \cite{tHooft:1977nqb}, which essentially carries magnetic charge in the fundamental weight lattice.  Such 't Hooft magnetic line operators have the defining property that they do not commute with Wilson loops.  This is to be contrasted with the GNO magnetic line operators relevant here, which commute with Wilson loops. However, in the popular modern approach of Kapustin \cite{Kapustin:2014gua} both types of magnetic line operators are referred to as 't Hooft line operators.  Note that $\Xi_a$ are ``genuine line operators'' since they require no topological surface to be gauge-invariant.  We will refer to these operators as  GNO 't Hooft magnetic holonomies, or ``magnetic holonomies" for short.

Thus, as far as magnetic charge is concerned, the monopole operators behave as if they arise from dimensional reduction of GNO 't Hooft magnetic line operators wrapping  $S^1$.  So the various $\theta$ vacua are labeled by expectation values of these magnetic operators, $\langle\Xi_a\rangle$.  But we have not yet commented on the electric charges of these operators.   In fact, we have often referred to the associated field configurations as `magnetic monopoles', and if one naively classifies field configurations as purely electric, purely magnetic, or dyonic, one might be  tempted to infer that monopole-instanton operators have vanishing electric charge and thus descend from a purely magnetic holonomy. This is not quite right, because the 3D monopole operators are not electric charge eigenstates, and so their electric charge is not well defined.  Below we explore this using Poisson duality, following an earlier analysis with Poppitz \cite{Poppitz:2011wy}.

To see the issues in the simplest context, consider the case $N=2$, and parametrize the holonomy $\Omega$ as
\begin{equation}
\Omega=\left(\begin{array}{cc}
e^{i \phi/2}\\
 & e^{-i \phi/2}
\end{array}\right)
\end{equation}
so that (in an obvious gauge-fixed sense)  $\left\langle A_{4}\right\rangle =\frac{1}{L}\frac{\sigma^{3}}{2}\phi$. The center-symmetric case corresponds to $\phi=\pi$. For $N=2$, there are two types of elementary monopole-instantons which give rise to the potential of Eq.~\eqref{eq:YM_potential}. It is well-known that these monopole-instanton solutions are self-dual and satisfy the BPS equations $E_{\mu}=\pm B_{\mu}$, see e.g.~Ref.~\cite{Davies:1999uw}. However, in the dimensionally reduced theory, the $A_4$ field becomes a scalar and acquires a vacuum expectation value rendering it massive.  If we make the dependence of the monopole-instanton operators on $\langle A_4 \rangle$ explicit, they take the form 
\begin{equation}\label{eq:mi_full_amp}
\mathcal{M}_{1}\sim e^{-\frac{8\pi}{\lambda} \phi}e^{+i\sigma},\qquad\mathcal{M}_{2} \sim e^{-\frac{8\pi}{\lambda}\left(2\pi-\phi\right)}e^{-i\sigma}
\end{equation}
to leading order in the semiclassical expansion and neglecting $\theta$ dependence for the time being.   The two solutions associated with $\mathcal{M}_{1,2}$ have the minimal values of the action, but monopole-instantons with higher action also exist.  These more general  solutions can be constructed by allowing generic  winding numbers for the compact scalar field $\phi$ \cite{Anber:2011de},
\begin{equation}
\phi(x_4=L)= \phi(x_4=0) + 2\pi n_W\quad\Leftrightarrow\quad\mathcal{M}(n_W)=e^{-\frac{8\pi}{\lambda}\left|\phi+2\pi n_{W}\right|}e^{-i\sigma} \, ,
\end{equation}
and $\mathcal{M}_1$ and  $\overline{\mathcal{M}}_2$ correspond to the cases $n_{W}=0$ and $n_{W}=-1$, respectively. 

Although the monopole-instantons are defined by a self-duality relation, \eqref{eq:mi_full_amp} shows why they should \emph{not} be interpreted as a dyonic monopoles \cite{Poppitz:2012sw,Anber:2017rch}. The monopole amplitude is of the form $e^{-\frac{4\pi}{g^2} \phi +i\sigma}$, rather than the dyonic form $e^{iq_{e} \phi+iq_{m}\sigma}$.  In particular,  $\phi$ exchange between monopole-instantons configurations of the same $\phi$-charges leads to an attractive force, since $\phi$ is a (massive) scalar.   This is not the correct sign for  electric interactions, and means that it is incorrect to interpret monopole-instantons as dyonic field configurations.

However, there is a heuristic but highly suggestive way to interpret the potential of \eqref{eq:YM_potential} as coming from an infinite sum over dyonic operators, as shown in \cite{Poppitz:2012sw,Poppitz:2011wy}.  To do so, consider the potential generated by a sum of the monopole-instantons with \emph{all} possible windings, $n_{W}$:   
\begin{equation}
\label{eq:all_winding_pot}
V_{\textrm{all windings} } = m_W^3\left[e^{i\sigma} \sum_{n_{w}\in\mathbb{Z}}e^{-\frac{8\pi}{\lambda}\left| \phi+2\pi n_{W}\right|} + \mathrm{h.\,c.}\right] + \ldots \;.
\end{equation}
The $\ldots$ in this expression represent the contributions from correlated events which start to contribute from $\mathcal{O}(e^{-2S_0})$, as well as the $\phi$ dependence generated in perturbation theory around each monopole-instanton. Since we have been considering the leading-order winding solutions all along, the differences between \eqref{eq:YM_potential} and \eqref{eq:all_winding_pot} are exponentially suppressed. The value of focusing on the terms shown explicitly in \eqref{eq:all_winding_pot} is that they have a well-defined and highly suggestive four-dimensional interpretation. Specifically,  as a consequence of Poisson resummation identities, the following relation holds \cite{Poppitz:2012sw,Poppitz:2011wy}
\begin{equation}\label{eq:n2_sum}
\left[\sum_{n_{w}\in\mathbb{Z}}e^{-\frac{8\pi}{\lambda}\left|\phi+2\pi n_{W}\right|}\right]e^{i\sigma}=\left[\frac{1}{\pi}\sum_{n_{e}\in\mathbb{Z}}\frac{\frac{8\pi}{\lambda}}{\left(\frac{8\pi}{\lambda}\right)^{2}+n_{e}^{2}}e^{in_{e} \phi}\right]e^{i\sigma}.
\end{equation}
While each of the terms on the left-hand side originates from field configurations with well-defined magnetic charge, all of them have an ill-defined electric charge.
\footnote{Relatedly, the individual terms on the left-hand side of \eqref{eq:n2_sum} are not periodic under $\phi \to \phi + 2\pi$, and this would not be cured by including the perturbative or correlated-event contributions we have been dropping.  But the $\phi$ periodicity must be respected by the full theory and taking into account the sum over $n_w$ cures this issue.  See \cite{Dorey:1999sj} for a discussion of this issue in the context of $\mathcal{N}=1$ super-Yang-Mills theory, where the extra constraints  provided by supersymmetry allow one to establish very explicit results in this direction.}   
But each of the terms on the right-hand side is of the form $e^{i n_e \phi + i \sigma}$, and hence \emph{can} be interpreted as originating from field configurations with well-defined electric and (GNO) magnetic charges $(q_e, q_m)=(n_e, 1)$. (Note that since we have specizlied to $N=2$ there is only one type of magnetic and electric charge.) 

When $\lambda$ is small, the sum on the left is dominated by the $n_W = 0, -1$ terms, corresponding to the $\mathcal{M}_1,\mathcal{M}_2$ operators, while the sum on the right is dominated by terms with charge $|n_e|\lesssim 8\pi/\lambda$. So for small $NL\Lambda$, one can interpret the monopole-instantons as arising from  a purely magnetic holonomy as well as many dyonic holonomies constructed from some combinations of the electric and magnetic holonomies.  The relevant dyonic-instanton configurations arise from dimensional reduction of the corresponding dyonic field configurations wrapping the $S^1$ direction.  Here the relevant dyonic line operators  can be constructed from the Wilson line with charge $\Omega\sim\left( 1,0\right)$  and the GNO t' Hooft line with charge $\Xi\sim\left( 0,1\right)$ as $\Omega^{q_e}\Xi^{q_m}\sim\left( q_e, q_m \right)$. 

Just as the holonomy acquiring a vacuum expectation value implies a well-defined vacuum expectation value of the dimensionally reduced $A_4$ scalar, we can view the $\sigma_k$ (the specialization of $\vec{\sigma}_k$ for $N=2$) as arising from a vacuum expectation value of the magnetic holonomy. Hence the specification of the vacuum structure of center-stabilized YM amounts to a specification of the ordinary and magnetic holonomy vacuum expectation values.

Generalizing to arbitray $N$, we simply require $N-1$ labels for all the GNO 't Hooft lines.
\footnote{In the special case of a center-symmetric theory considered here, all magnetic holonomies take on the same vacuum expectation value so it may seem redundant to use $N-1$ labels. But if center symmetry is spontaneously broken, then the expectation values of magnetic holonomies associated to distinct simple roots of $su(N)$ would not have to be the same, so we keep the labels distinct. } 
The sum of \eqref{eq:n2_sum} becomes
\begin{equation}\label{eq:arbn_sum}
\sum_{a=1}^N\left[\sum_{n_{w}\in\mathbb{Z}}e^{-\frac{8\pi}{\lambda}\left|\vec{\alpha}_a \cdot\vec{\phi}+2\pi n_{W}\right|}\right]e^{i\vec{\alpha}_a \cdot\vec{\sigma}}=\sum_{a=1}^N\left[\frac{1}{\pi}\sum_{n_{e}\in\mathbb{Z}}\frac{\frac{8\pi}{\lambda}}{\left(\frac{8\pi}{\lambda}\right)^{2}+n_{e}^{2}}e^{in_{e}\vec{\alpha}_a \cdot\vec{\phi}}\right]e^{i\vec{\alpha}_a \cdot\vec{\sigma}}.
\end{equation}
By the same reasoning above, the vacuum expectation values of $\phi^a$  and $\vec{\sigma}_k$  can equally well be described as vacuum expectation values of the Polyakov loop and magnetic holonomy $\left\langle\Xi_{a}\right\rangle$. In particular, as already noted above and illustrated in Fig.~\ref{fig:vacuaYM}, on the $k$-th $\theta$ extremum, the magnetic holonomy has the expectation value
\begin{align}
e^{i \vec{\alpha}_a \cdot \vec{\sigma}} \bigg|_{\langle \vec{\sigma} \rangle = (2\pi k/N) \vec{\rho} }= e^{2\pi i k/N}\,.
\end{align}

The fact that the confining phase at small $NL\Lambda$ is characterized by vanishing expectation values for Wilson holonomies and non-vanishing expectation values for GNO 't Hooft holonomies fits in a satisfying way with the standard expectations about the phases of gauge theories.  Confinement is supposed to be associated with an area law for large Wilson loops, which is indeed observed in the $NL\Lambda$ regime   \cite{Unsal:2008ch,Poppitz:2017ivi}.  The area law behavior is also associated with the absence of a disconnected contribution in the holonomy correlator
\begin{align}
\langle \tr \Omega(\vec{x}) \tr \Omega(0) \rangle = \langle \tr \Omega(\vec{x}) \tr \Omega(0) \rangle_{\rm connected} \equiv e^{-\beta E(|\vec{x}|)},
\end{align}
so that the Polyakov loop expectation value vanishes \cite{Polyakov:1978vu}, and $E(|\vec{x}|) = \mathfrak{\sigma}_{\rm tension} |\vec{x}|$ for large $|\vec{x}|$.  At the same time, large GNO 't Hooft loops have a perimeter law.  One can verify that this is indeed the case in center-stabilized YM theory due to the Coulomb interactions between the dual photons.  And not coincidentally, we have seen that certain GNO 't Hooft holonomy operators have non-vanishing expectation values, so that GNO 't Hooft holonomy correlation functions have non-vanishing disconnected pieces.

Lastly, it is interesting to recall that when $\theta\to\theta+2\pi$, the charges of dyonic operators transform as  $(q_e, q_m) \longrightarrow ( q_{e}+q_{m},q_{m})$ \cite{Witten:1979ey}.  In the context of \eqref{eq:arbn_sum} this amounts to a transformation
\begin{equation}
e^{in_{e}\vec{\alpha}_a\cdot\vec{\phi}}e^{i\vec{\alpha}_a\cdot\vec{\sigma}}\to e^{i\left(n_{e}+1\right)\vec{\alpha}_a\cdot\vec{\phi}}e^{i\vec{\alpha}_a\cdot\vec{\sigma}}=e^{i2\pi/N}e^{in_{e}\vec{\alpha}_a\cdot\vec{\phi}}e^{i\vec{\alpha}_a\cdot\vec{\sigma}}.
\end{equation}
In view of these remarks, one one can interpret the $e^{2\pi i /N}$ jumps in the phase of the magnetic-monopoles (and associated dyonic line operators) as a function of $\theta$ as arising from the Witten effect.

\subsection{Branch dependence of gluon condensates}
\label{sec:condensates}

We have seen above that as $\theta$ varies, there are phase transitions between $\theta$ vacua, and one of the order parameters for these phase transitions is the expectation value of the GNO 't Hooft magnetic holonomies on $S^1$.  These phase transitions are also marked by non-analyticities in the expectation values of \emph{local} operators.  Here we illustrate this point by calculating the parity-even and parity-odd gluon condensates  $\langle \frac{1}{N}\tr F_{\mu \nu} F_{\mu \nu} \rangle $ and $\langle \frac{1}{N}\tr F_{\mu \nu} \tilde F_{\mu \nu} \rangle $.
\footnote{We thank Tin Sulejmanpasic for discussions of the statistical  method of evaluating condensates in the semiclassical domain which we use below. This method  was  used in  the context of ${\cal N}=1$ SYM to show vanishing of the gluon condensate in \cite{Behtash:2015kna}.  In purely bosonic theory, an alternative approach to the computation of $\langle \tfrac{1}{N} \tr F \tilde{F} \rangle$, with the same result, is described in \cite{Bhoonah:2014gpa}.  }

The leading contribution to these condensates in the calculable $NL\Lambda \ll 1$ domain arises from monopole-instantons.  Before explaining quantitatively, we note that this statement is in sharp distinction with old QCD literature which gives the impression that vacuum condensates arise from 4d instantons.  The important distinction between these two cases is that the dilute monopole-instanton gas on $\mathbb R^3 \times S^1$  is a controlled semi-classical approximation to the Euclidean vacuum of the gauge theory, while the ``dilute instanton gas" is not a controlled approximation. Therefore, it should not be a surprise that monopole-instantons produce the correct multi-branched structure of observables, while 4d instantons do not.  Relatedly,  the monopole-instanton density is of order $e^{-S_0}$, while 4d instanton density is order $e^{-N S_0}$.  In the large $N$ limit the 4D instanton contribution to the vacuum energy is exponentially suppressed, while the monopole-instanton contribution is not suppressed.  As a result,  the $\theta$ angle dependence  induced by the monopole instantons is consistent with  large-$N$ arguments \cite{Witten:1980sp}, in contrast to the $\theta$-dependence inferred from naive instanton calculations.

At leading order in semi-classics, we can think of the vacuum as a dilute gas of $N$ types of monopole-instantons (in a grand canonical ensemble), each with complex fugacity  
\begin{align}
z_a =  e^{-S_0 + i \theta/N},\qquad a = 1, \ldots, N \, .
\end{align}
Physically, the fugacity of any given species of monopole-instanton is
$
|z_a| \sim \mathcal{N}_{{\cal M}_a} / V_{\mathbb R^3 \times S^1}
$, 
where $\mathcal{N}_{{\cal M}_a}$ is the average (with respect to the path integral measure) number of monopole-instantons of type $a$ in the spacetime volume $V_{\mathbb{R}^3 \times S^1}$. 
In this statistical interpretation, we can write
\begin{align} 
 \langle \tr F_{\mu \nu} F^{\mu \nu} \rangle  \sim 
 \frac{ \int_{\mathbb R^3 \times S^1}  \tr F_{\mu \nu} F^{\mu \nu} }{V_{\mathbb{R}^3 \times S^1} } \, . 
\end{align}
That is, the value of the gluon condensate can be obtained by computing the average value of $  \tr F_{\mu \nu} F^{\mu \nu} $ in a spacetime volume $V_{\mathbb R^3 \times S^1}$.   In terms of the fugacities defined above, to leading order in the semiclassical expansion, this implies that
\begin{align}
\langle \tr F_{\mu \nu} F_{\mu \nu} \rangle  \sim \frac{8\pi^2}{N} \times \frac{m_W^3}{L} \times \sum_{a=1}^{N} \left(z_{a} + z_{a}^{\dag} \right) \,.
\end{align}
where $\frac{8\pi^2}{N}$ is the value of $\int d^4x \tr F_{\mu\nu} F^{\mu \nu}$ on any given monopole-instanton, $\frac{m_W^3}{L} $ is the relevant inverse volume, and the sum is over the $N$ species of monopole-instantons, with $z_{a}^{\dag}$ the fugacities of the anti-monopole-instantons.  As a result, to leading order in the semiclassical expansion at small $NL\Lambda$, we obtain
\begin{align}
 \langle \frac{1}{N}\tr F_{\mu \nu} F_{\mu \nu} \rangle  
 &\sim   m_W^4  e^{-S_0} \cos\left(\frac{2\pi k+\theta}{N}\right) \bigg|_{k = k^{*}(\theta)}\nonumber  \\
 &= 
 \Lambda^4  
(\Lambda L N)^{-1/3}  \cos\left(\frac{2\pi k+\theta}{N}\right)  \bigg|_{k = k^{*}(\theta)}
\label{p-even}
\end{align} 
where $k^{*}(\theta)$ is the value of $k$ that maximizes $\cos\left(\frac{2\pi k+\theta}{N}\right) $ for any given $\theta$.  Setting $k = k^{*}(\theta)$ ensures that we evaluate  $\langle \frac{1}{N}\tr F_{\mu \nu} F_{\mu \nu} \rangle$ on the vacuum branch, which minimizes the energy,  for any given value of $\theta$. 

 \begin{figure}[t]
\begin{center}
\includegraphics[width=0.45\textwidth]{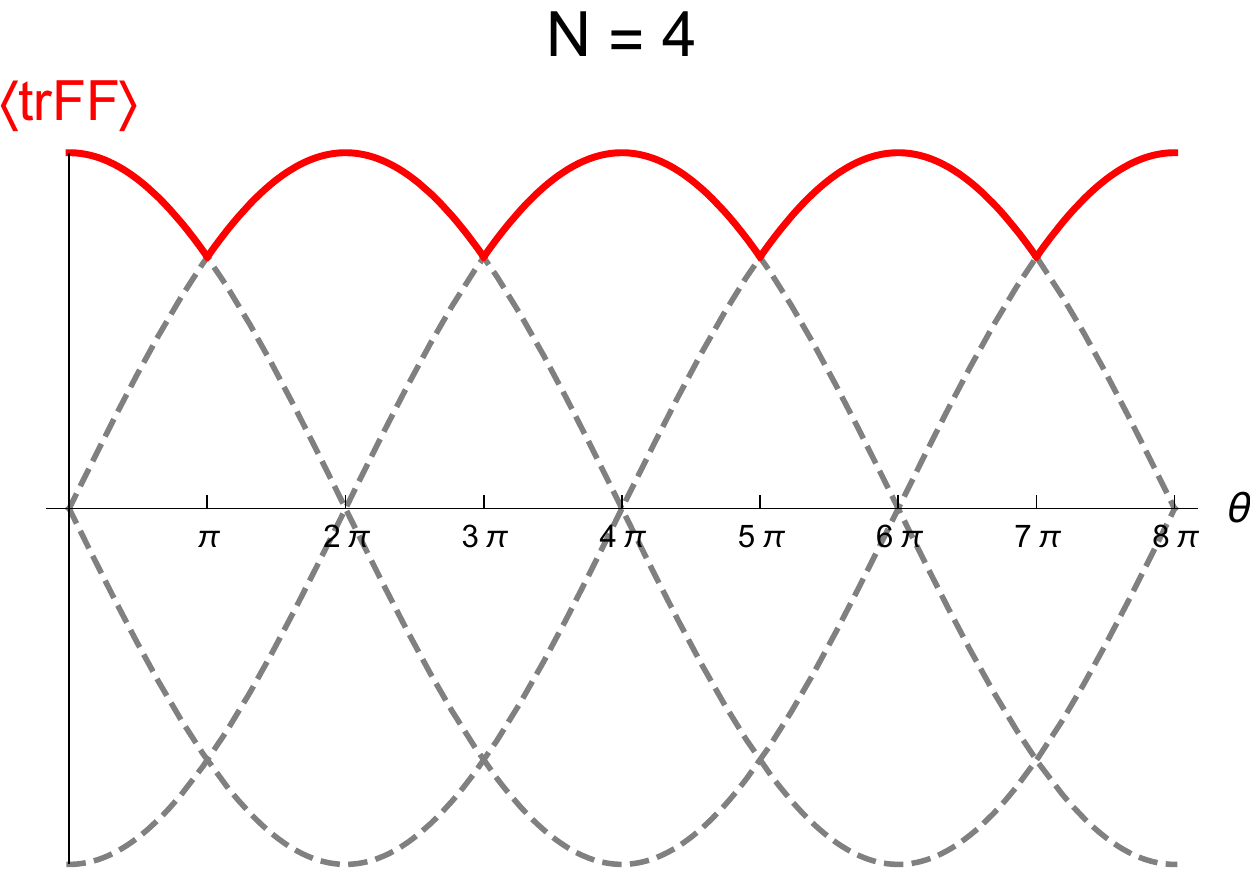}
\includegraphics[width=0.45\textwidth]{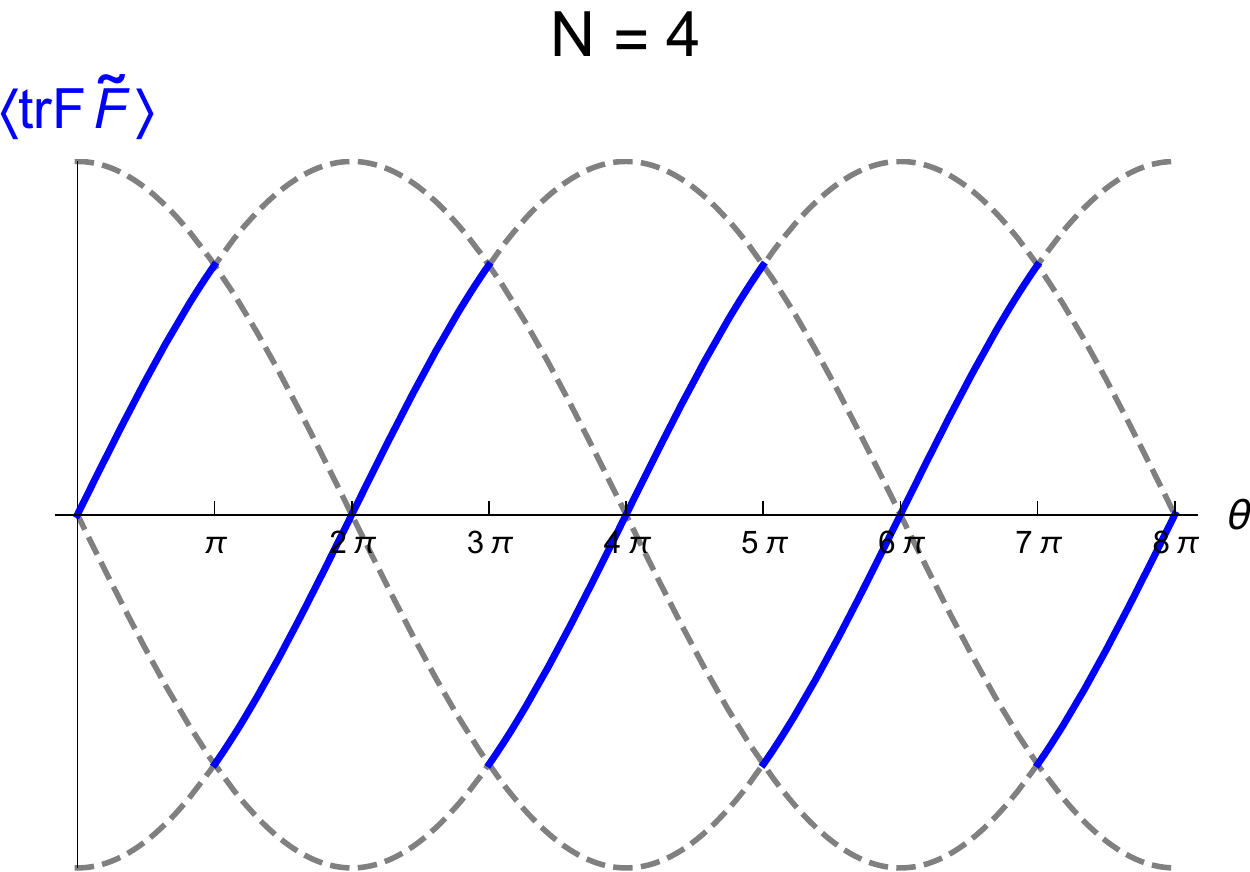}
\caption{[Color Online.] A sketch of the $\theta$ angle dependence of  the lowest-dimension non-trivial parity even and odd condensate $ \langle \frac{1}{N}\tr F_{\mu \nu} F_{\mu \nu} \rangle(\theta)$ and   $ \langle \frac{1}{N}\tr F_{\mu \nu} \widetilde F_{\mu \nu} \rangle(\theta)$  for $N=4$ in $SU(N)$ YM theory in the semiclassical confining domain.  The continuous bold red curve shows the values of the parity-even condensate, while the saw-tooth bold blue curves denote the values of the parity-odd condensate.  The grey dashed curves mark the values of the condensates on the $N$ different $\theta$ extrema.   }
\label {fig:condensate-Gr}
\end{center}
\end{figure}

The discussion for the parity odd gluon condensate is almost the same except for the fact that 
$\int_{\rm lump}  \tr F_{\mu \nu} \widetilde F_{\mu \nu}  = \pm \int_{\rm lump}  \tr F_{\mu \nu}   F_{\mu \nu} $, with a plus sign  for monopoles and a minus sign for antimonopoles.   Hence, the anti-monopole contribution comes with a negative over-all sign, leading to 
\begin{align}
 \langle \frac{1}{N}\tr F_{\mu \nu}  \widetilde F_{\mu \nu} \rangle   &\sim  m_W^4  e^{-S_0} \sin \left(\frac{2\pi k+\theta}{N}\right)\bigg|_{k = k^{*}(\theta)} \nonumber  \\
 &= 
  \Lambda^4  
(\Lambda L N)^{-1/3}  \sin \left(\frac{2\pi k+\theta}{N}\right)  \bigg|_{k = k^{*}(\theta)} 
\label{p-odd}
\end{align} 
Note that we wrote \eqref{p-odd} as an operator statement in Minkowski space.  In Euclidean space, there is an extra factor of $i$ for operators like $\tr F_{\mu \nu}  \widetilde F_{\mu \nu} = \tr \epsilon^{\mu \nu \alpha \beta} F_{\mu \nu}   F_{\alpha \beta} $  that involve the spacetime Levi-Civita tensor.

In the decompactification limit where  $LN\Lambda \gg 1$, we expect these semiclassical results to approach
\begin{align}
 \langle \tfrac{1}{N}\tr F_{\mu \nu} F_{\mu \nu} \rangle  &\sim  \Lambda^4    f_{\rm even}\left(\frac{2\pi k+\theta}{N}\right)\bigg|_{k = k*(\theta)}   \\
  \langle \tfrac{1}{N}\tr F_{\mu \nu}  \widetilde{F}_{\mu \nu} \rangle   &\sim  
     \Lambda^4 \,   g_{\rm odd} \left(\frac{2\pi k+\theta}{N}\right)\bigg|_{k = k*(\theta)}  
\label{eq:condenates}
\end{align} 
where $f_{\rm even}(x)$ and $g_{\rm odd}(x)$ are even and odd $2\pi$-periodic functions.

Note that 
$ \langle \frac{1}{N}\tr F_{\mu \nu} F_{\mu \nu} \rangle$ is a continuous function of $\theta$ with a cusp at $\theta=\pi$. 
The largest deviation of the observable from its value at $\theta=0$ takes place at $\theta=\pi$, and the difference is $O(1/N^2)$. Hence, at $N=\infty$, this condensate is actually a constant, independent of $\theta$ \cite{Unsal:2012zj}.  On the other hand, the parity odd condensate
$ \langle \frac{1}{N}\tr F_{\mu \nu} \widetilde F_{\mu \nu} \rangle$ is a saw-tooth function of $\theta$  with a discontinuity at $\theta=\pi$. 
The largest deviation of the observable from its value at $\theta=0$ takes place at $\theta=\pi$,  where it is given by 
$\pm \Lambda^4 \sin \left(\frac{\pi} {N}\right) $.  Since the parity-odd condensate is zero at  $\theta=0$, and it is  $\mathcal{O}(N^{-1})$ at $\theta=\pi$, at $N=\infty$, it  must remain zero for all values of the $\theta$-angle.  
This means that   spontaneous  $T$ (or $CP$) -breaking is not visible at leading order in the large $N$ expansion, and only becomes apparent once $1/N$ corrections are taken into account.

\section{Genericity of spinodal points in the YM phase diagram}
\label{sec:genericity}

In this section we first show that when $NL\Lambda$ is small, center-stabilized YM has spinodal curves in its $\theta$-$NL\Lambda$ phase diagram.  A spinodal point is defined as a place where a metastable vacuum reaches its limits of stability, and the theory becomes exactly ``gapless" on the metastable branch, in the sense that eigenvalues of the Hessian matrix of the potential around the vacuum vanish.    A natural follow-up question is whether the spinodal curves persist all the way to large $NL\Lambda$, where the physics approaches that of pure YM theory on $\mathbb{R}^4$.  We then seek to gather some intuition on this question by examining the $\theta$-dependence of various related theories:  QCD(adj) at small $NL\Lambda$ \cite{Unsal:2007jx}, and QCD with light fundamental fermions on both $\mathbb{R}^4$ and $\mathbb{R}^3\times S^1$.  

\subsection{YM on $\mathbb{R}^3\times S^1$}

\begin{figure}[t]
\begin{center}
\includegraphics[width=0.7\textwidth]{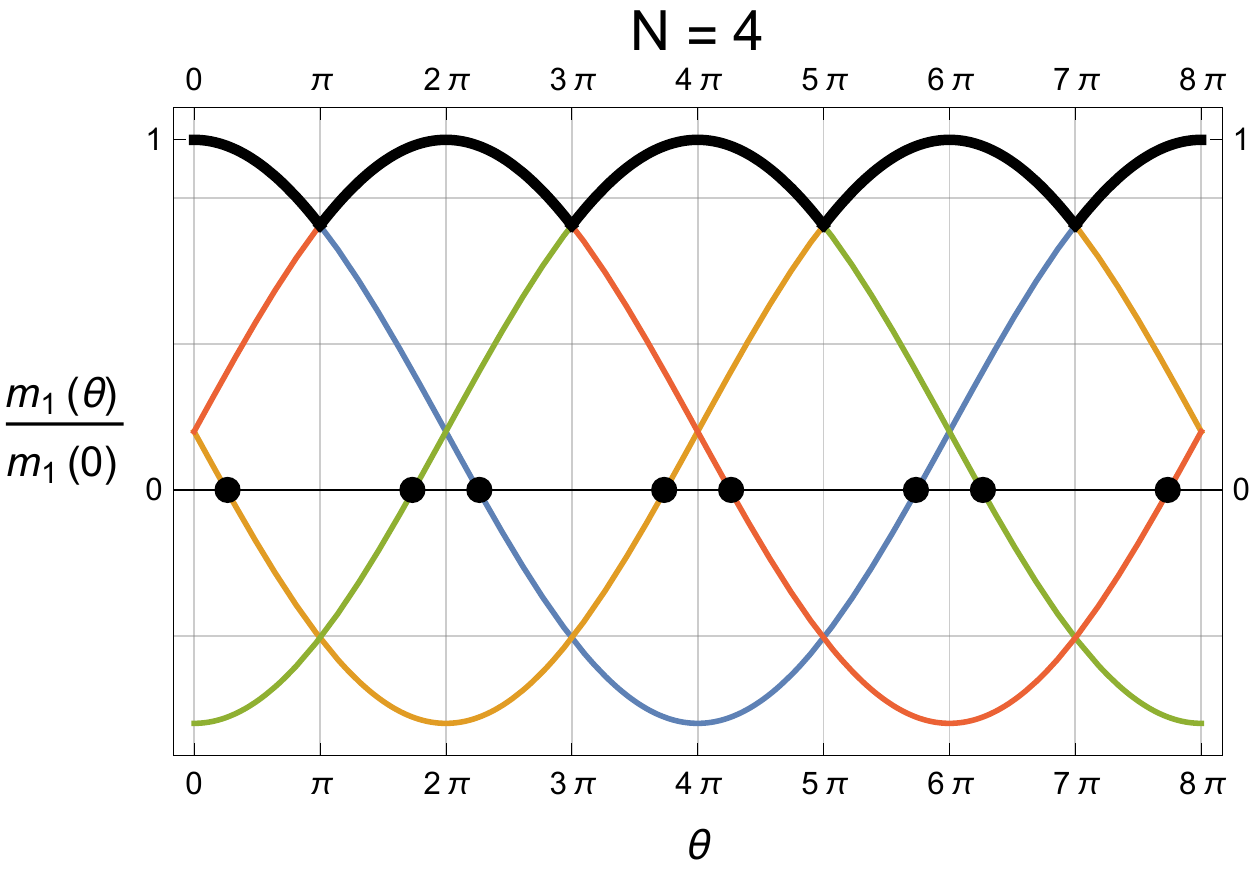}
\caption{[Color Online.]  Plot of the mass of the lightest  excitation of YM  theory with $N=4$  at small $NL\Lambda$ as a function of $\theta$, normalized to its value at $\theta = 0$, with magnetic bion effects taken into account.  The thick black curve at the top marks the behavior of the gap along the thermodynamically stable branch.  Each of the four colored curves show the dependence of the mass on $\theta$ in a given $k$-extremum, with blue, red, green, orange corresponding to $k=0,3,2,1$ respectively.  The black dots mark spinodal points, and the fact that there are two distinct spinodal points near $\theta = 2\pi k, k \in \mathbb{Z}$ is due to magnetic bions.     }
\label {fig:spinodal4}
\end{center}
\end{figure}

Suppose we start at $\theta = 0$ in the $k=0$ global minimum configuration, and adiabatically increase $\theta$,  the system will stay in the $k=0$ state so long as it is locally stable.  The time scale for bubble nucleation is parametrically long so long as $NL\Lambda \ll 1$ \cite{Bhoonah:2014gpa,Li:2014lza}.%
\footnote{For example, if one takes the large $N$ limit with $NL\Lambda$ held fixed and small, and $\theta > \pi$, the decay rate is $\Gamma \sim e^{- \mathfrak{C}\, N^{7/2}/(\pi - \theta)^2}$ where $\mathfrak{C} \sim e^{+S_0/2}$\cite{Li:2014lza}.}   
But the $k=0$ extremum becomes locally unstable for $\theta > N\pi/2 + \mathcal{O}( e^{-S_0} )$.  This means that at $\theta = \theta^*$
\begin{align}
\theta^* = N\pi/2 + c\, e^{-S_0}   , \qquad c>0
\end{align}
there is a spinodal point. Here, $c\sim\mathcal{O}(1)$ to leading log accuracy. The positivity of $c$ is tied to the positivity of the bion contributions to the effective potential, and their independence from $\theta$.  Consequently one can balance a negative leading-order contribution to the masses against positive contributions from higher orders by tuning $\theta$, and thus arrange for the dual photon ``masses" from Eq.~\eqref{eq:extremum_spectrum} to vanish for $k=0$ at some $\theta$. So, if $\theta$ is adiabatically increased from $0$ to $\theta^*$, the system will become ``gapless", in the sense that the eigenvalues of the Hessian of the potential evaluated on the $\theta = 0$ vacuum configuration vanish at $\theta^{*}$.   The existence of these spinodal points is thus a robust feature of YM theory with stabilized center symmetry at small $L$. 
The spinodal points of YM theory with $N=4$  at small $NL\Lambda$ are illustrated in Fig.~\ref{fig:spinodal4}.

 \begin{figure}[t]
\begin{center}
\includegraphics[width=0.7\textwidth]{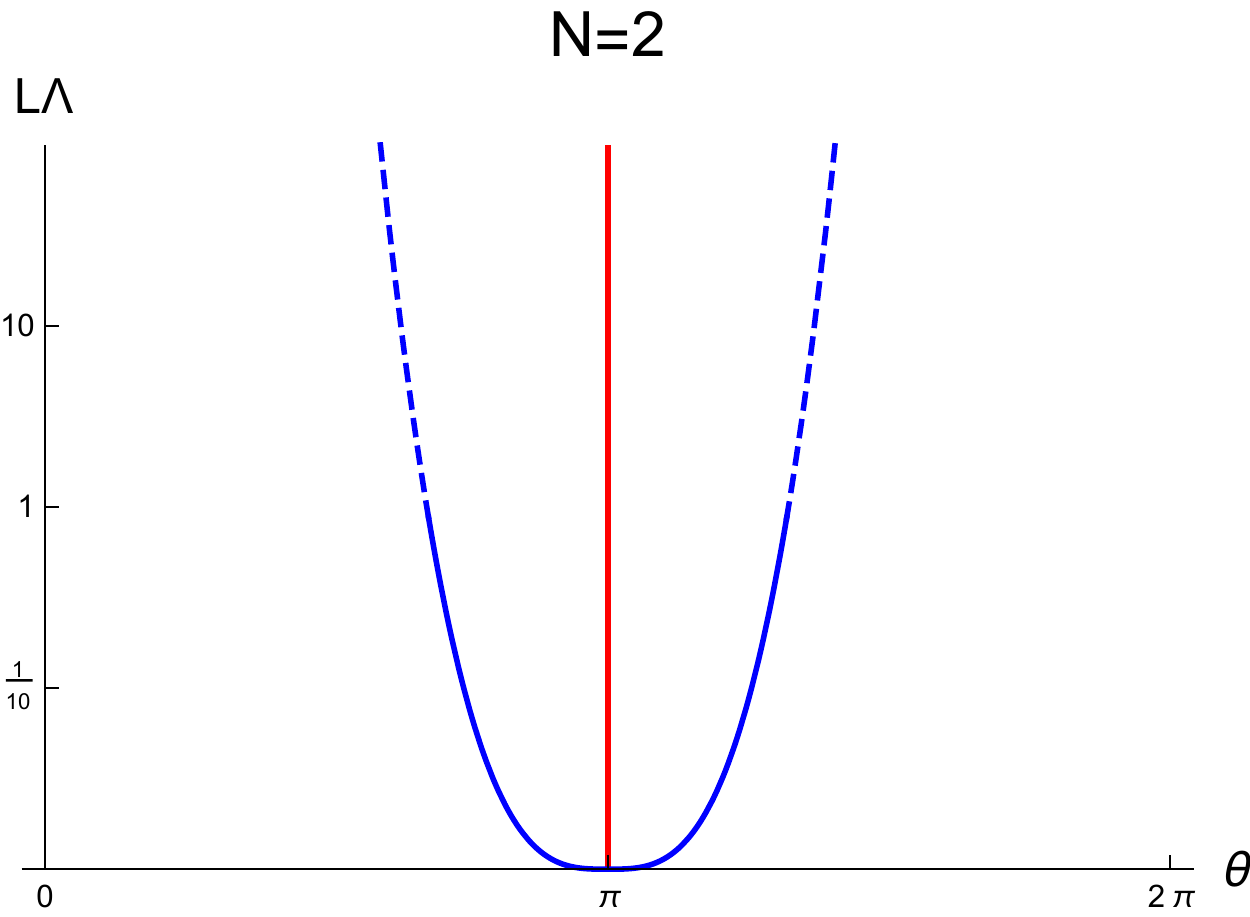}
\caption{[Color Online.] Spinodal curves (blue) and phase transition lines (vertical red line) for center-stabilized $SU(2)$ YM theory in the $L\Lambda$-$\theta$ plane.  The vertical red line at $\theta = \pi$ indicates a first-order phase transition between two $\theta$-vacua.  The two blue curves mark the regions of spinodal instability of the $\theta$-vacua, where a metastable branch reaches the limit of local stability.  The behavior of the spinodal curves for $L\Lambda \lesssim 1$ follows from  \eqref{eq:spinodal_theta_0} and \eqref{eq:spinodal_theta_2pi}, and hence is indicated by a solid curve, while the behavior for $L\Lambda \gtrsim 1$ is a conjecture and is indicated by the dashed portion of the dashed curve.   }
\label {fig:spinodal_curves}
\end{center}
\end{figure}

To illustrate these features in the simplest context, consider $N=2$.  The spinodal curves for $N=2$ are sketched in Fig.~\ref{fig:spinodal_curves}.   In terms of the  physical dual photon field $\alpha_1 \cdot \vec{\sigma} \equiv \sigma$, the potential associated with the $k = 0,1$ extrema can be written as 
\begin{align}
V_{k} = - \frac{A}{\lambda^{2}}m_{W}^{3}e^{-S_0} \left[2\cos\left(\sigma\right)\cos\left( \frac{\theta+2\pi k}{2}\right)  +  c\, e^{-S_0} \cos(2\sigma) + \ldots \right].
\label{eq:two_color_potential}
\end{align}
The magnetic bion $\mathcal{M}_1 \overline{\mathcal{M}}_2$ contribution to the effective potential is never smaller in absolute value than the topological bion $\mathcal{M}_1\mathcal{M}_1$ contribution \cite{Unsal:2012zj}, so in writing \eqref{eq:two_color_potential} we have dropped the topological bion contributions.
  
To leading order, when $\theta$ is  smaller than $\pi$ the $k=0$ branch is locally stable while the $k=1$ branch is locally unstable.   If $\theta$ becomes  larger than $\pi$ the branches exchange roles.  Since for any given $\theta$ there is only one locally stable state, it is also the globally stable state. At $\theta = \pi$ the leading-order contributions to the $\sigma$ mass vanish.   (In fact here the leading-order potential itself vanishes, but this is special to $N=2$.)  But the magnetic bion contribution to the effective masses is always positive.  As a result, if we start in the stable $k=0$ vacuum at $\theta = 0$ and adiabatically increase $\theta$, the theory has a spinodal point at
\begin{align}
\theta^{*}_{+} = \pi + 4\,c \, e^{-S_0}  +\mathcal{O}\left(e^{-2S_0}\right) \,.
\label{eq:spinodal_theta_0}
\end{align}
If we instead start at $\theta = 2\pi$, where the stable vacuum has $k = 1$ and adiabatically decrease $\theta$, there will be a spinodal point at
\begin{align}
\theta^{*}_{-} = \pi - 4\,c \, e^{-S_0}  +\mathcal{O}\left(e^{-2S_0}\right) \,.
\label{eq:spinodal_theta_2pi}
\end{align}
The phase diagram of the theory is sketched in Fig.~\ref{fig:spinodal_curves}.

\subsection{SYM and QCD(adj)}
It is known that ${\cal N}=1$ SYM and QCD(adj) with exactly massless fermions $\lambda_A, A = 1, \ldots, n_{\rm adj}$ have unbroken center symmetry  on $\mathbb R^3 \times S^1$ for any $S^1$ size provided that the fermions are endowed with periodic boundary conditions.
\footnote{For adjoint QCD with $n_{\rm adj}>1$ flavors of adjoint Majorana, strictly speaking this is a conjecture, which is however strongly supported by all available analytic and numerical lattice simulation evidence.  For $n_{\rm adj} = 1$, the preservation of center symmetry can be proved using a combination of semiclassical analysis along with holomorphy arguments enabled by the emergence of supersymmetry for this number of fermion flavors, since adjoint QCD with $n_{\rm adj} =1$ is precisely $\mathcal{N}=1$ SYM theory.} 
In the chiral limit, where the fermions are exactly massless, the $\theta$ angle can be eliminated (``rotated away") using field redefinitions in the path integral, and the physics is $\theta$-independent. Turning on a small common mass term  $m_{\rm adj}$ for the adjoint fermions brings $\theta$ dependence back into the physics.  Working with a small $S^1$ and varying $m_{\rm adj}$ we thus obtain an interesting setting where the $\theta$ dependence of the physics becomes calculable. Our emphasis here is the local stability vs. instability of vacua labeled by $k$.

It is known that in QCD(adj) as well as SYM with  $m_{\rm adj} =0$, the mass gap for gauge fluctuations and linear confinement is caused by magnetic bion mechanism \cite{Unsal:2007jx}.  The magnetic bions produce $\sigma$ interaction terms of the form
\begin{align}
e^{-2 S_0} e^{ i (\vec{\alpha}_{a} - \vec{\alpha}_{a+1})\cdot\vec{\sigma}  } \,,
\end{align}   
At the same time, when $m_{\rm adj} =0$, monopole-instantons have $2n_{\rm adj}$ fermionic zero modes and thus cannot contribute to bosonic potentials, so the magnetic bions give the dominant contribution to the $\sigma$ potential.  The monopole-instantons induce Yukawa-type interactions for $\vec{\sigma}$ with the Cartan components of the adjoint fermions $\vec{\lambda}_{A}$.  These interaction terms take the form
\begin{align}
e^{-S_0} e^{ i \vec{\alpha}_{a}\cdot\vec{\sigma}  }  \prod_{A=1}^{n_{\rm adj}}( \vec{\alpha}_{a}\cdot\vec{\lambda}_{A} )^{2}\,,\qquad m_{\rm adj} = 0
\end{align}
  Consequently, the monopole-instantons do not induce a potential for dual photons.  However, turning on a small mass 
for fermions, the fermi zero modes are ``soaked up" by the mass term, and as a result the monopole-instanton generated interactions take the form
\begin{align}
e^{-S_0} \left(\frac{m_{\rm adj}}{m_W}\right)^{n_{\rm adj}}  e^{ i \vec{\alpha}_{a}\cdot\vec{\sigma}  + i \theta/N } \,.
\end{align}
 As long as $(m_{\rm adj}/m_W)^{n_{\rm adj}} \lesssim  e^{-S_0}$,  the bions provide the dominant mechanism of confinement and mass gap,  while once $(m_{\rm adj}/m_W)^{n_{\rm adj}} \gtrsim  e^{-S_0}$ the physics becomes dominated by monopole-instantons.\footnote{An exception is $SU(2)$ YM theory at $\theta = \pi$, because there the monopole-instanton-induced potential cancels exactly even for finite $m_{\rm adj}$.}

 \begin{figure}[t]
\begin{center}
\includegraphics[width=0.7\textwidth]{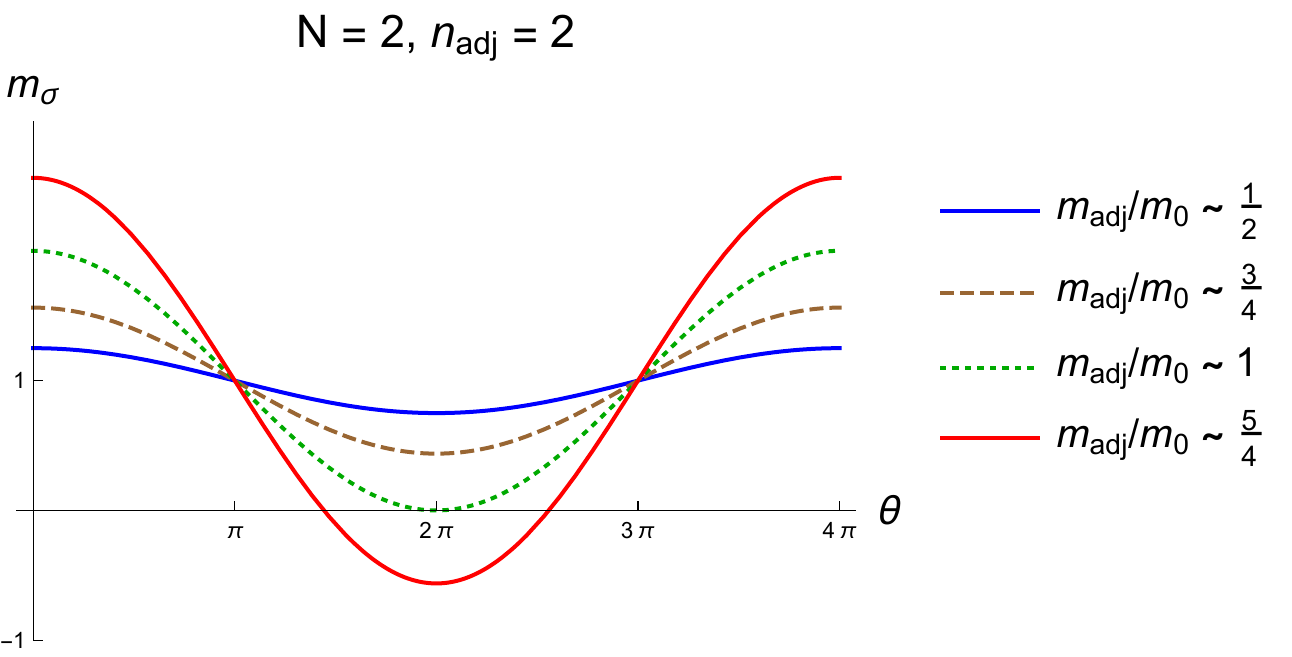}
\caption{[Color Online.] A sketch of the $\theta$ angle dependence of the dual photon mass $m_{\sigma}$ in $SU(2)$ adjoint QCD with $n_{\rm adj}=2$ flavors of adjoint Majorana fermions in the $k=0$ $\theta$-vacuum, as a function of the fermion mass $m_{\rm adj}$ in units of the mass gap at vanishing adjoint quark mass, $m_0 \sim m_W e^{-S_0}$.  
For small enough fermion mass,  the $\theta$-dependence is very mild, in the sense that the $k=0$ $\theta$-vacuum is stable for all $\theta$.   The dual photon mass is normalized to its value in the chiral limit. But once $m_{\rm adj}$ exceeds $m_0$, the $\theta$-dependence becomes more dramatic:  there is a region where the  $k=0$ vacuum becomes locally unstable.    }
\label {fig:vacuumAdj}
\end{center}
\end{figure}

This has an interesting implication in light of the discussion in the previous section. In the regime where the mass gap in $k$-vacua is induced by bions, it is fairly easy to show that there is no local instability of any branch.  In fact, the branch dependence of the vacua is a tiny perturbation 
which has no effect on their stability.   All $N$ branches are  stable.  To see this, first note that the non-perturbatively induced potential is  
\begin{equation}
V\left(\vec{\sigma}\right)= - m_W^3 \left[\frac{m_{\rm adj}}{m_W}\right]^{n_{\rm adj}}  e^{-S_0}\sum_{a=1}^{N}\cos\left[\vec{\alpha}_{a}\cdot\vec{\sigma}+\frac{\theta}{N}\right] -  e^{-2S_0} \sum_{a=1}^{N} \cos\left[ (\vec{\alpha}_{a} -\vec{\alpha}_{a+1})\cdot \vec{\sigma} \right] 
\end{equation}
The first and second terms above come from the first and second orders in the semiclassical expansion, respectively.  However, in the  ${(m_{\rm adj}})^{n_f}  \ll e^{-S_0}$ regime, the latter term is actually parametrically larger and dominates the dynamics. (This does not invalidate the expansion.) Diagonalizing the mass matrix, one finds 
\begin{equation}
m_{q,k}^2 \sim m_W^2 \left[ \left(\frac{m_{\rm adj}}{m_W}\right)^{n_{\rm adj}}   e^{-S_0} \sin^{2}\left(\frac{\pi q}{N}\right)\cos\left(\frac{2\pi k+\theta}{N}\right) + e^{-2S_0}   \sin^{4}\left(\frac{\pi q}{N}\right) \right]
\label{eq:extremum_spectrum_qcd}
\end{equation} 
and   in the $(m_{\rm adj}/m_W)^{n_{\rm adj}} \ll e^{-S_0}$ regime, the mass gaps for all branches are positive definite.  In the chiral limit, the mass gap is of course $\theta$-independent.  This can be traced to the fact that magnetic bions give the dominant contribution to the bosonic potential near the chiral limit, and they carry zero topological charge and hence do not bring in any $\theta$ dependence.

For $n_{\rm adj} =1$,
the ${\cal N}=1$ SYM case,  there is an analytically calculable center symmetry  changing phase transition for    
$m_{\rm adj} \sim m_W e^{-S_0}$ \cite{Poppitz:2012sw,Poppitz:2012nz} and in order to see the effect on branches, one needs to add  a center-stabilizing double-trace deformations.\footnote{Intuitively, this is because in the supersymmetric theory the center-stabilizing effective potential is non-perturbatively small, while for $n_{\rm adj}\ge 2$ the holonomy effective potential is non-zero and stabilizes center symmetry already at one loop.}  However, when $1 < n_{\rm adj} < 5.5$, we can   increase  $m_{\rm adj}$ until $m_{\rm adj} \sim m_W$ without any obstacles.  As a result, when  $n_{\rm adj} \geq 1$, we observe that some of the branches start to become unstable around $m_{\rm adj}/m_W \sim e^{-S_0/n_{\rm adj}}$ while retaining control over the long-distance dynamics.  In the  $e^{-S_0} \ll (m_{\rm adj}/m_W)^{n_{\rm adj}} \ll 1 $ regime, the behavior of the branches coincides precisely with the one we discussed in the context of center-stabilized YM theory.  This is of course not an accident:  $n_{\rm adj}$ massive fermions with $m_{\rm adj} \lesssim m_W$ are the prototypical example of a center-stabilizing deformation of Yang-Mills theory.  The evolution of the local stability of the $k$-vacua are illustrated in  Fig.~\ref{fig:vacuumAdj}.

\subsection{QCD with light fundamental quarks}

 \begin{figure}[t]
\begin{center}
\includegraphics[width=0.7\textwidth]{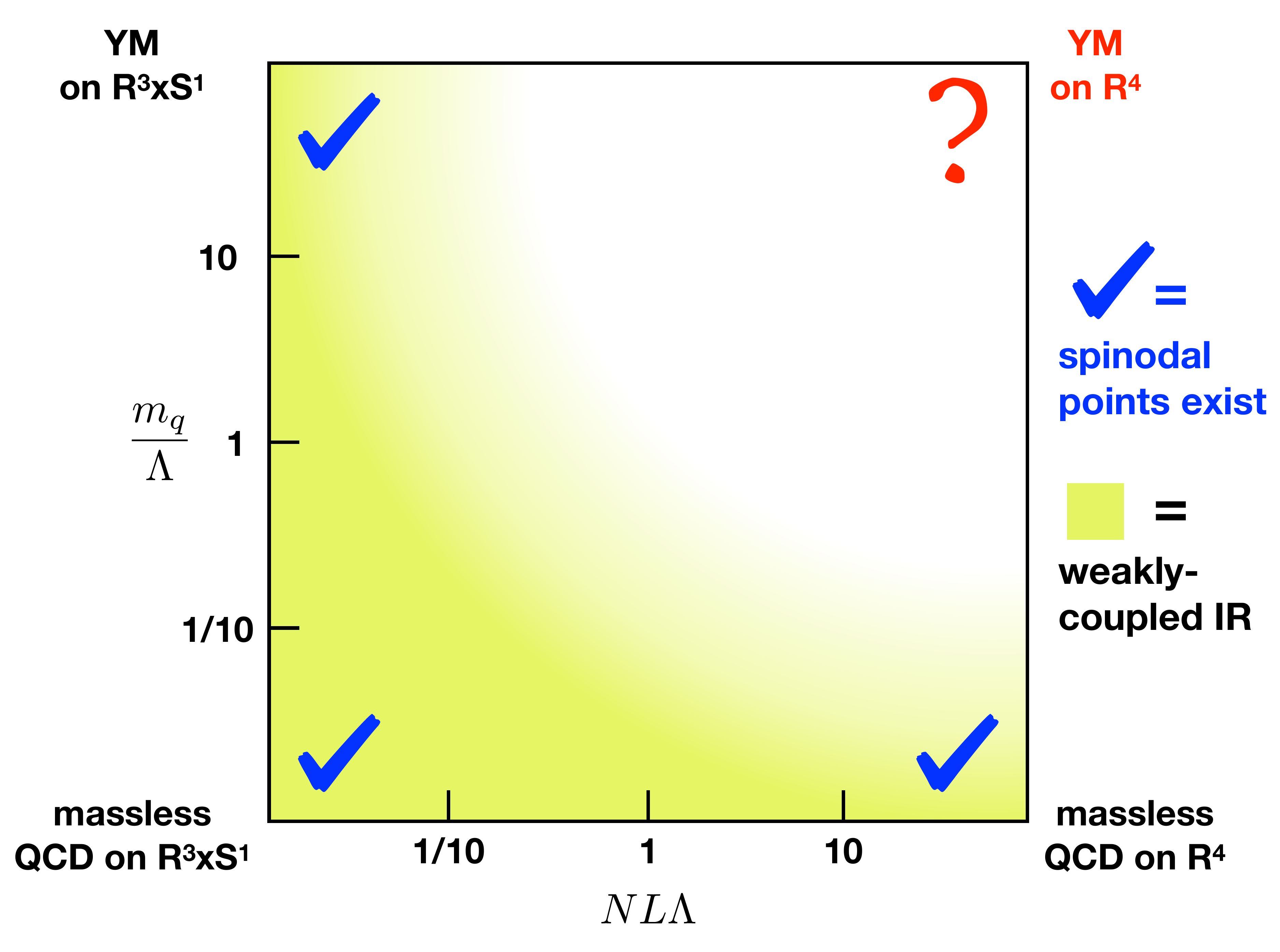}
\caption{[Color Online.]   A cartoon mapping out the regions in gauge-theory parameter space where spinodal points exist.  We consider QCD on $\mathbb{R}^3\times S^1$ with boundary conditions that preserve a $\mathbb{Z}_N$ color-flavor-center (CFC) symmetry \cite{Cherman:2017tey}, and vary the quark mass $m_q$ and the circle size $L$, assuming that CFC symmetry is preserved at small $L$.  When $m_q\gg \Lambda$ the CFC symmetry reduces to the usual center symmetry.   In all regions where the physics can be quantitatively explored (shaded yellow), there are spinodal points which appear as a function of  $\theta$.}
\label {fig:spinodal_square}
\end{center}
\end{figure}

We now consider QCD with massive fundamental quarks and look for spinodal points as a function of $\theta$.  The value of considering this example is that when the quarks are light, there is an approximate spontaneously broken chiral symmetry on $\mathbb{R}^4$.    The low energy physics is then systematically describable using chiral effective field theory, and thus one can probe the existence of spinodal points directly on $\mathbb{R}^4$.  The chiral perturbation theory results described in this section have been known for a long time, see e.g. \cite{Creutz:1995wf,Smilga:1998dh,Dubovsky:2010je}.  The only novelty has to do with the relation to center symmetry  described below.

Take QCD with $N_F = N$ quarks with a common quark mass $m_q$ and a theta angle $\theta$.  Then, if the theory is placed on  $\mathbb{R}^3 \times S^1_L$, one can pick the quark boundary conditions in such a way that the theory has an exact $\mathbb{Z}_N$ color-flavor-center (CFC) symmetry \cite{Cherman:2017tey}, see also \cite{Kouno:2012zz,
    Sakai:2012ika,
    Kouno:2013zr,
    Kouno:2013mma,
    Iritani:2015ara,
    Kouno:2015sja,
    Hirakida:2016rqd,
    Hirakida:2017bye,
    Sulejmanpasic:2016uwq,
    Cherman:2016vpt}. 
This recently uncovered symmetry of QCD combines color center symmetry transformations with cyclic flavor permutations.   By itself, color center symmetry is broken in the presence of fundamental quarks.  But a diagonal combination of color center and cyclic flavor permutations is a bona fide symmetry of the theory, with order parameters which depend non-trivially on the parameters of the system.  

Our motivation in considering this example is two-fold.  First, it has an especially smooth connection to pure Yang-Mills theory:  there is a center-type 0-form $\mathbb{Z}_N$ symmetry for all values of $m_q$, which at small $m_q$ must be thought of as the CFC symmetry described above, while at large $m_q$ it can just as well be thought of as the 0-form part of $\mathbb{Z}_N$ center symmetry of pure YM theory.  We expect the existence of this symmetry to be important for studying the $\theta$ dependence,  because of the key role center symmetry plays in our Yang-Mills analysis in the preceding section.  Second, if one ensures that CFC symmetry does not break for any $S^1$ size $L$, for instance by a double-trace deformation, then chiral symmetry will continue to be broken at  both large and small $L$ \cite{Cherman:2016hcd,Cherman:2017dwt}, and must be broken for all $L$ if CFC symmetry is unbroken as shown in \cite{Cherman:2017dwt} from discrete anomaly-matching considerations.  

Let us work at very large $L$, where for our immediate purposes we can neglect the twisted boundary conditions for the quarks.  (Their main effect is simply that the charged Nambu-Goldstone bosons get effective masses $\sim 1/(LN)$.)   The leading-order terms in the chiral Lagrangian are
\begin{align}
\mathcal{L}_{\chi PT} = f_{\pi}^2 \tr \partial_{\mu} U \partial^{\mu} U + B \tr( M U + M^{\dag} U)
\end{align}
Here $M = m_q \mathbf{1}_{N_F}$ is the quark mass spurion field, while $U$ is the $N_F \times N_F$ unitary chiral field, whose exponent contains the Nambu-Goldstone fields.   We will take $m_q$ complex with $\arg m_q = \theta/ N_F$, which is equivalent to turning on a $\theta$ angle in the QCD Lagrangian with the conventional normalization.

To analyze the vacuum structure, let us focus on diagonal matrices $U$.  This amounts to setting to zero the charged pseudo-Nambu-Goldstone fields.  This is motivated by the expectation that the vector-like part of the flavor symmetry will not break spontaneously.  Moreover, as we will see, the issue of spinodal instabilities will simply come down to the possibility that the neutral pion mass-squares can change sign at tree level, as a function of $\theta$.  But at the quadratic level, the charged fields cannot mix with neutral fields.

For simplicity, let us illustrate the story by considering the case $N_F = N = 2$.\footnote{For $N=2$ the fundamental representation is pseudoreal, so the chiral symmetry is enhanced from $SU(2)\times SU(2)$ to $SU(4)$.  The $SU(4)$ symmetry is spontaneously broken  to $ Sp(4)$.  Consequently, there are five Goldstone bosons:  the familiar $\pi^{\pm}, \pi^0$, and two light "di-quark" baryons $d^{\pm}$. Since we are only consider neutral field configurations, the $d^{\pm}$ and 
$\pi^{\pm}$ do not affect our analysis.}  Focusing on neutral field configurations amounts to setting 
\begin{align}
U\big|_{\rm diag} = e^{i \pi^0 \tau_3 /f_{\pi}} \,.
\end{align}
In terms of a dimensionless pion field, $\sigma = \pi^{0}/f_{\pi}$, the potential term in the chiral Lagrangian is just
\begin{align}
V(\sigma) = 2 B m_q \left[\cos(\sigma+\theta/2)+ \cos(-\sigma+\theta/2)\right] + \cdots
\label{eq:chiPT_V}
\end{align}
where $\cdots$ stand for higher-order terms in the chiral Lagrangian, such as e.g. $\tr M^2 U^2$.

With the chosen notation, this expression, which arises from chiral perturbation theory on $\mathbb{R}^4$, is clearly exactly of the same form we obtained for two-color YM theory at small $NL\Lambda$ in \eqref{eq:YM_potential}.  The only difference is in the prefactor of the potential, which depends on $m_q$ in QCD with light quarks on $\mathbb{R}^4$, but becomes $m_q$-independent for large $m_q$.    

Lastly, for small-$(NL\Lambda)$, \eqref{eq:chiPT_V} also coincides with the chiral Lagrangian derived in \cite{Cherman:2016hcd}.   Consequently, once bion effects/higher-order chiral expansion corrections  effects, we again find spinodal behavior as a function of $\theta$.

In all of these examples, which have a $\mathbb{Z}_2$ center-like symmetry, there are two $\theta$ ``vacua'', but at $\theta = 0$, only one of them is stable.  The other is locally unstable.  Precisely at $\theta = \pi$, the two vacua switch roles, and at $\theta = \pi$, time-reversal symmetry is spontaneously broken. But, if $\theta$ is changed adiabatically, one can stay on a metastable branch as one passes through $\theta$.  Eventually this metastable branch reaches a limit of stability, which is a spinodal point where the $\pi^0$  becomes ``gapless".  With light quarks, this always happens due to a balance between the contributions to the $\pi^0$ mass term from the leading-order chiral Lagrangian against contributions from a higher order.  At small $L$ in center-stabilized YM theory, the same thing happens due to a balance between monopole-instanton contributions and the magnetic bion contributions.  The situation is summarized in Fig.~\ref{fig:spinodal_square}.

Of course, it is an open question whether spinodal points as a function of $\theta$ also exist in pure YM theory, which corresponds to the question marks in the upper-right corner of the diagram in Fig.~\ref{fig:spinodal_square}.  But the fact that the phenomenon is present everywhere one can compute in Fig.~\ref{fig:spinodal_square}, as well as directly on $\mathbb{R}^4$ in a holographic model of YM theory, makes it  plausible  that the occurrence of spinodal points as a function of $\theta$ may be a general feature of $SU(N)$ gauge theory with exact $\mathbb{Z}_N$ center or color-flavor-center symmetry, so long as there are no light  adjoint-representation fermions.

Finally, we note that we assumed that $N$ is finite in the discussion above.  In the large $N$ limit with fixed $N_F$, the $\eta'$ meson becomes light, and the potential in the chiral Lagrangian becomes
\begin{align}
V =  B \tr( M U + M^{\dag} U) + a (\log \det U)^2 \, ,
\end{align}
where $B \sim \mathcal{O}(N^1), a \sim \mathcal{O}(N^0)$, to leading order in the $1/N$ and $m_q / f_{\pi}$ expansions.  Restricting $U$ to the Cartan subgroup, one can then verify that the conditions for a vanishing set of first derivatives and vanishing second derivatives cannot be simultaneously satisfied for any $\theta$ and $a \neq 0$.  This means that a given $\theta$ extremum cannot go from being locally stable to being locally unstable, and thus spinodal behavior is not possible in the large $N$ limit of QCD with light fundamental quarks.  But of course, this is not so relevant to the question of whether spinodal points can emerge when the fundamental quarks are heavy.  

\section{Summary}
\label{sec:outlook}
We have studied the vacuum structure of $SU(N)$ YM theory as a function of $\theta$, using compactification on $\mathbb{R}^3\times S^1$ with stabilized center symmetry as an arena to explore the dynamics.  Our results fit nicely with conventional wisdom concerning the $\theta$ dependence, in that as expected, observables are $N$-branched functions of $\theta$ due to the existence of many candidate $\theta$  vacua.  However, we find that for any given $\theta$ there are only $\approx N/2$ candidate vacua which are locally stable.  As $\theta$ is varied, some candidate vacua cease to exist as locally-stable field configurations, but new locally-stable candidate vacua appear, and eventually take their turn being global minima. We discuss the physical interpretation of the $\theta$ vacua, and find that they can be distinguished by the expectation values of certain magnetic line operators that carry GNO charge but not 't Hooft charge.  Finally, as a corrolary to some of the results above,  we find that YM theory has spinodal points as a function of $\theta$, at least in the domain of validity of our analysis.  It may be interesting to explore whether the presence of such spinodal points might have some phenomenological applications in e.g. axion model building.


\section*{Acknowledgments}

We are grateful to  P.~Argyres, A.~Kapustin, M.~Shifman, T.~Sulejmanpasic, and M.~Yamazaki for helpful conversations, and owe a special thanks to L.~G.~Yaffe for insightful comments and advice on a preliminary version of the manuscript.   K.~A. is supported by the U.S.~Department of Energy under Grant No.~DE-SC0011637. A.~C. and M.~\"U.  thank the KITP for its warm hospitality as part of the program `Resurgent Asymptotics in Physics and Mathematics' during the final stages of the research in this paper. Research at KITP is supported by the National Science Foundation under Grant No. NSF PHY11-25915.  A.~C. is also supported by the U. S. Department of Energy via grants DE-FG02-00ER-41132, while M.~\"U. is supported U. S. Department of Energy grant DE-FG02-03ER41260. 


\newpage

\bibliographystyle{JHEP}
\bibliography{small_circle}

\end{document}